\documentclass[a4paper,fleqn,usenatbib,useAMS]{mnras}

\usepackage{graphicx}	
\usepackage{amsmath}	
\usepackage{amssymb}	
\usepackage{multicol}        
\usepackage{bm}		
\usepackage{pdflscape}	
\usepackage[encapsulated]{CJK}
\usepackage{ucs}
\usepackage[utf8x]{inputenc}
\usepackage{xcolor}

\usepackage[T1]{fontenc}
\usepackage{ae,aecompl}

\usepackage{newtxtext,newtxmath}

\title[Rings and arcs around evolved stars (II)]{
  Rings and arcs around evolved stars. II.
  The Carbon Star AFGL\,3068 and the Planetary Nebulae NGC\,6543,
  NGC\,7009 and NGC\,7027}
\author[Guerrero et al.]{
M.A.\ Guerrero$^{1}$\thanks{E-mail:\,mar@iaa.es},
G.\ Ramos-Larios$^{2}$,
J.A.\ Toal\'{a}$^{3}$,
B.\ Balick$^4$,
and L.\ Sabin$^5$ \\
$^{1}$Instituto de Astrof\'{i}sica de Andaluc\'{i}a (IAA-CSIC), Glorieta de la Astronom\'{i}a S/N, 18008 Granada, Spain \\
$^{2}$Instituto de Astronom\'{i}a y Meteorolog\'{i}a, Universidad de Guadalajara, Av. Vallarta 2602, Arcos Vallarta, 44130 Guadalajara, Mexico\\
$^{3}$Instituto de Radioastronom\'{i}a y Astrof\'{i}sica (IRyA), UNAM Campus Morelia, Apartado postal 3-72, 58090 Morelia, Michoacan, Mexico \\
$^4$Department of Astronomy, University of Washington, Seattle, WA 98195-1580,
USA \\
$^5$Instituto de Astronom\'{i}a, Universidad Nacional Aut\'onoma de M\'exico,
Apdo.\ Postal 877, 22860 Ensenada, B. C., Mexico
}

\pubyear{2018}

\begin{document}
\label{firstpage}
\pagerange{\pageref{firstpage}--\pageref{lastpage}}
\maketitle

\begin{abstract}

We present a detailed comparative study of the arcs and fragmented ring-like
features in the haloes of the planetary nebulae (PNe) NGC\,6543, NGC\,7009,
and NGC\,7027 and the spiral pattern around the carbon star AFGL\,3068 using
high-quality multi-epoch \emph{HST} images.
This comparison allows us to investigate the connection and possible 
evolution between the regular patterns surrounding AGB stars and the 
irregular concentric patterns around PNe.
The radial proper motion of these features, $\simeq$15 km~s$^{-1}$, 
are found to be consistent with the AGB wind and their linear sizes
and inter-lapse times ($500-1900$ yr) also agree with those found
around AGB stars, suggesting a common origin.
We find evidence using radiative-hydrodynamic simulations that regular
patterns produced at the end of the AGB phase become highly distorted by 
their interactions with the expanding PN and the anisotropic illumination
and ionization patterns caused by shadow instabilities.   
These processes will disrupt the regular (mostly spiral) patterns
around AGB stars, plausibly becoming the arcs and fragmented rings
observed in the haloes of PNe.

\end{abstract}




\section{Introduction}
\label{sec:intro}

Back in the mid 80s, planetary nebulae (PNe) were still considered to be
single shells of ionized material resulting from a massive episode of
mass-loss marking the end of the asymptotic giant branch (AGB) phase (the
so-called superwind) during the late evolution of low- and intermediate-mass
(1 M$_{\odot} \lesssim$ M$_\mathrm{i} \lesssim$8 M$_{\odot}$) stars.  
The discovery of extended haloes, low surface-brightness
detached shells, and the identification of attached shells
surrounding the bright main nebulae
\citep{CJA1987,FBR1990,Betal1992} revealed them as
multi-layered shells of gas, each shell providing us with
information on different episodes of mass-loss in the late 
evolution of AGB stars \citep{ST1995}.

Accordingly, a PN can consist of different shells, including from
inside out a hot bubble and bright nebular rim, a nebular envelope and
a halo.  
The nebular envelope is attributed to the superwind, 
a dense, slow, and most-likely neutral wind, which produces 
dusty shells that enshroud the central star \citep[e.g.,][and 
references therein]{Cox2012,Mauron2013}.  
As the post-AGB star increases its temperature and subsequently its ionizing
flux, the nebular shell is rapidly (or nearly instantly) photo-ionized,
producing a sharp increase of the thermal pressure inside.
At the same time, the post-AGB star develops a fast stellar wind that
pushes outwards on the nebular envelope, producing a sharp rim of
swept-up material, whereas the fast stellar wind is shock-heated and
reaches X-ray-emitting temperatures, producing a hot bubble that can
be detected through its diffuse X-ray emission
\citep[e.g.,][]{Chu2001,Guerrero2002,Kastner2001}.  
Because of the dynamical effects of these processes, the 
signatures from the previous mass loss history are mostly 
erased.  
Finally, the haloes are attributed to the last thermal pulse experienced 
by the AGB star before leaving this phase \citep{ST1995,GM1999}.

\begin{table*}
\centering
\setlength{\columnwidth}{0.2\columnwidth}
\setlength{\tabcolsep}{1.30\tabcolsep}
\caption{General information of the sources in our sample}
\begin{tabular}{lcccrl}
\hline
\multicolumn{1}{l}{Source} & 
\multicolumn{1}{c}{RA} & 
\multicolumn{1}{c}{DEC} & 
\multicolumn{1}{c}{Distance} &
\multicolumn{1}{c}{T$_\mathrm{eff}$} & 
\multicolumn{1}{c}{Spectral Type} \\ 
\multicolumn{1}{c}{} & 
\multicolumn{2}{c}{(J2000)} & 
\multicolumn{1}{c}{(kpc)} & 
\multicolumn{1}{c}{(K)} & 
\multicolumn{1}{c}{} \\ 
\hline
AFGL\,3068 &  23 19 12.6  &  $+$17 11 33.1 &  1.0           &   2200 & Carbon Star  \\
NGC\,6543  &  17 58 33.4  &  $+$66 37 58.7 &  1.63$\pm$0.19 &  63000 & Of-WR(H)     \\
NGC\,7009  &  21 04 10.8  &  $-$11 21 48.5 &  1.15$\pm$0.15 &  82000 & O(H)         \\
NGC\,7027  &  21 07 01.8  &  $+$42 14 10.0 &  1.0           & 215000 & $\dots$      \\
\hline
\end{tabular}
\vspace{0.4cm}
\label{sample}
\end{table*} 


In the late 90s, a new PN structural component was added to this 
list: concentric, ring-like features or arcs detected immediately 
outside their bright inner nebular shells.  
Actually, these ring-like structures were first reported in dust-scattered
\emph{Hubble Space Telescope (HST)} images of the proto-PNe (extended sources
just evolving into the PN stage) CRL\,2688, IRAS\,17150$-$3224, and
IRAS\,17441$-$2411 \citep{Kwok1998,Sahai1998,Su1998}, and immediately after
in the PNe Hubble\,5, NGC\,7027, and the very remarkable case of NGC\,6543
\citep[][]{Terzian2000,Balick2001}.
Later \emph{HST} and ground-based optical and space mid-infrared
\emph{Spitzer} images of emission lines from ionized gas or 
continuum emission scattered by dust have found an increasing number
of them \citep[e.g.][and references therein]{Corradi2004,RL2011,GM2015}.
In \citet[][hereafter Paper I]{RL2016}, ring-like features were searched
systematically in a sample of $\sim$650 PNe and proto-PNe with suitable 
\emph{HST} and \emph{Spitzer} archival images.
These features were found to be ubiquitous to all types of nebular
morphologies, but present only in a reduced $\simeq$8\% fraction
of sources.   
For consistency with Paper~I, we will generically refer to these 
arcs and fragmented rings as \emph{rings}, even though their shapes 
can only approximately and locally be fitted by concentric circles.

The large angular size and low-surface brightness of these ring-like
features have hampered their characterisation.  
\citet{Balick2001} presented a detailed analysis of the rings in NGC\,6543,
a.k.a.\ the Cat's Eye Nebula, using \emph{HST} images, concluding that they
have a thickness of $\sim$1000~AU and an ejection time-lapse of 1500~yr.
The optical and mid-infrared study of IC\,418 presented by \citet{RL2012}
confirmed ejection time lapses of the same order $\sim$630~yr for the three 
concentric rings surrounding the bright main nebula.  
The clear ring-like structures around CRL\,618 and CRL\,2688 offer 
a unique opportunity to investigate this phenomenon among proto-PNe. 
A comprehensive investigation of the formation of the proto-PN CRL\,2688
using multi-epoch \emph{HST} observations \citep{Balick2012} implied 
proper motions of 0\farcs011~yr$^{-1}$ for the rings, from which ejection 
time-lapses of 100~yr during the last 4000~yr with an almost constant
ejection velocity of 18~km~s$^{-1}$ are derived.  
A similar investigation, however, yielded inconclusive results for the 
rings around the proto-PN CRL\,618 \citep{Betal2013}.

\begin{table*}
\centering
\setlength{\columnwidth}{0.2\columnwidth}
\setlength{\tabcolsep}{1.30\tabcolsep}
\caption{Description of the \emph{HST} datasets used for analysis}
\begin{tabular}{llllcrrc}
\hline
\multicolumn{1}{l}{Source} & 
\multicolumn{1}{c}{Detector} & 
\multicolumn{1}{c}{Aperture} & 
\multicolumn{1}{c}{Filter} & 
\multicolumn{1}{c}{Date} & 
\multicolumn{1}{c}{Exposure time} & 
\multicolumn{1}{c}{Program} &
\multicolumn{1}{c}{Purpose} \\ 

\multicolumn{1}{c}{} & 
\multicolumn{1}{c}{} & 
\multicolumn{1}{c}{} & 
\multicolumn{1}{c}{} & 
\multicolumn{1}{c}{} & 
\multicolumn{1}{c}{(s)} & 
\multicolumn{1}{c}{} &
\multicolumn{1}{c}{} \\ 
\hline
AFGL\,3068 & ACS/WFC & WFC1 & F606W & 2004-09-18 & 1388~~~~ & 10185~~ & EXP        \\
AFGL\,3068 & ACS/WFC & WFC1 & F606W & 2010-10-20 & 9804~~~~ & 11676~~ & IMG,SB,EXP \\ 
NGC\,6543  & WFPC2   & PC1  & F502N & 1994-09-18 & 1600~~~~ &  5403~~ & EXP        \\
NGC\,6543  & WFPC2   & PC1  & F502N & 2000-09-15 & 1200~~~~ &  8390~~ & EXP        \\
NGC\,6543  & ACS/WFC & WFC  & F502N & 2002-05-04 &  700~~~~ &  9026~~ & IMG,SB     \\ 
NGC\,7009  & WFPC2   & WF3  & F502N & 2000-04-07 &  320~~~~ &  8114~~ & IMG        \\ 
NGC\,7027  & WFPC2   & PC1  & F555W & 1995-08-21 &  300~~~~ &  6119~~ & IMG        \\ 
NGC\,7027  & WFPC2   & PC1  & F555W & 2008-08-13 &  260~~~~ & 11122~~ & IMG,EXP    \\
    \hline
  \end{tabular}
\vspace{0.4cm}
\label{obs}
\end{table*}

The origin of ring-like structures in PNe and proto-PNe
is the result of sustained, thousand-year quasi-periodic
instabilities in the poorly understood AGB wind ejection
process.
Different scenarios have been proposed to address their formation, including 
i) viscous momentum (de)coupling between outflowing gas and dust
\citep[e.g.][]{Simis2001},
ii) solar-like magnetic inversions \citep[e.g.][]{GS2001},
iii) periastron passage of a stellar companion \citep[e.g.][]{Harpaz1997},
and
iv) variability in the wind of the AGB progenitor \citep[e.g.][]{Zijlstra2002},
to mention a few.
Still, our current knowledge of the formation and evolution of these
features among PNe and proto-PNe is not complete, but the answers 
should be looked for in the previous stage of the stellar evolution,
during the late AGB phase.

\begin{figure*}
\begin{center}
\includegraphics[width=\linewidth]{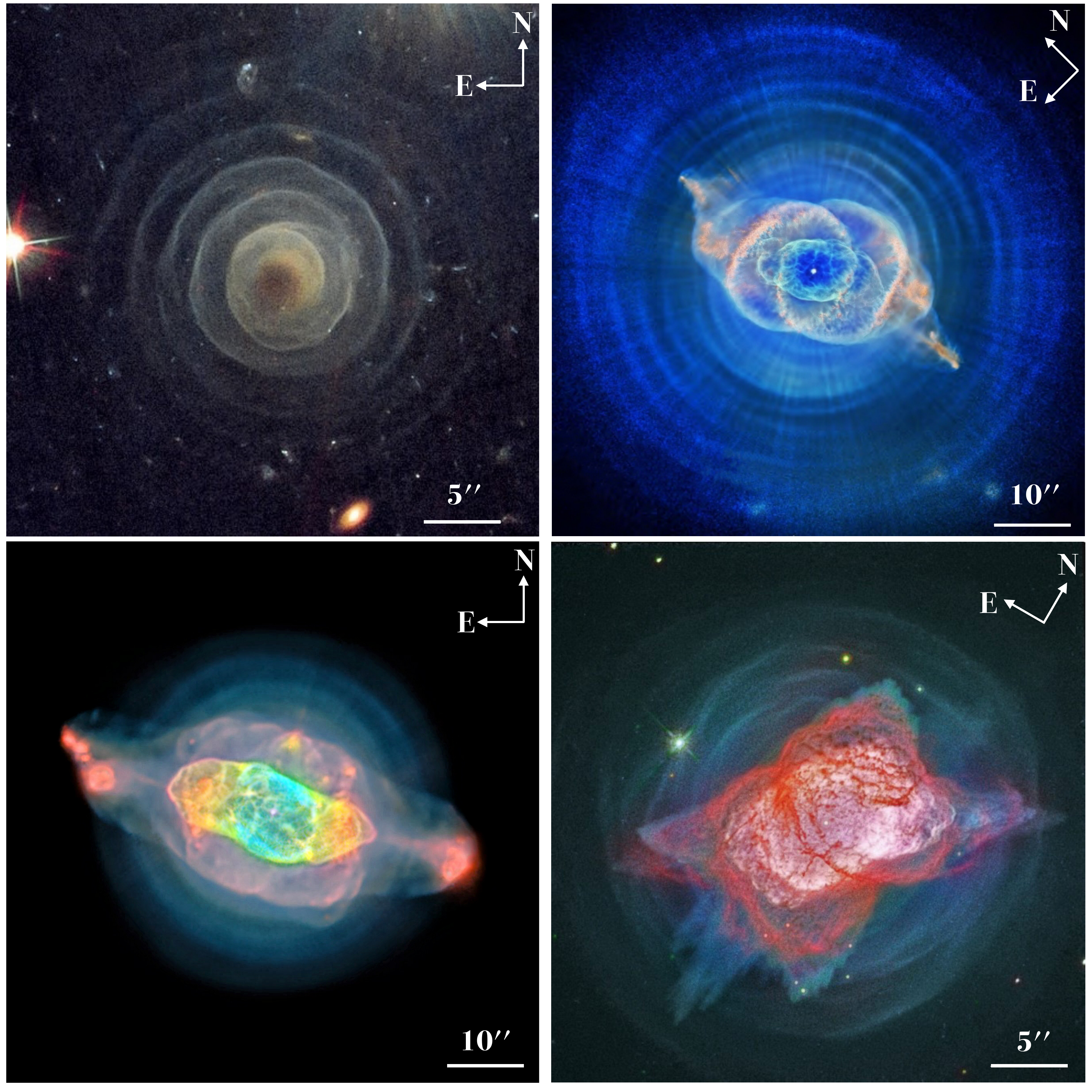}
\caption{
Colour-composite images of AFGL\,3068 (top left), NGC\,6543
(top right), NGC\,7009 (bottom left) and NGC\,7027 (bottom right).
The original images downloaded from the websites
https:{\slash\slash}www.flickr.com{\slash}photos{\slash}kevinmgill
(AFGL\,3068), 
http:{\slash\slash}hubblesite.org{\slash}image{\slash}1578{\slash}news$\_$release{\slash}2004-27 and 
https:{\slash\slash}apod.nasa.gov{\slash}apod{\slash}ap080322.html
(NGC\,6543), 
https:{\slash\slash}www.eso.org{\slash}public{\slash}spain{\slash}images{\slash}eso1731a and 
https:{\slash\slash}www.spacetelescope.org{\slash}images{\slash}opo9738g
(NGC\,7009), and 
https:{\slash\slash}apod.nasa.gov{\slash}apod{\slash}ap180109.html
(NGC\,7027) were combined and processed to enhance low surface-brightness
features.
These low surface-brightness external features are best seen in the
broad-band \emph{HST} F606W and F555W filters for AFGL\,3068 and
NGC\,7027, respectively, but in the narrow-band \emph{HST} F502N
filter for NGC\,6543 and NGC\,7009.  
}
\label{fig:pics}
\end{center}
\end{figure*}

Detailed mapping of the AGB circumstellar dust distribution in scattered
light at optical wavelengths has produced conspicuous results, including
the detection of the first detached shell and spiral structure around the
carbon-rich AGB stars TT\,Cyg \citep{Oetal1998} and AFGL\,3068
\citep{Mauron2006}, respectively.  
Subsequent sub-millimeter observations have detected
many other cases of regular patterns around AGB stars
\citep[e.g.,][]{Oetal1996,Mayer_etal2013,DL2009}
and even peered into the velocity field of amazing 3D 
spiral patterns
\citep[e.g.,][]{Maercker2012,Cernicharo2015,Kim2017}.
Currently, the most accepted scenario for the formation of spirals 
around AGB stars is that of binary interactions, where a binary 
companion in a highly eccentric orbit influences the spherical AGB
wind to imprint a spiral structure \citep{Mastrodemos1999}.  
More recent detailed hydrodynamic simulations support those early 
results 
\citep[][and references therein]{Kim2019}.

It is intriguing that the regular spiral patterns observed occasionally among
AGB stars are not detected around evolved sources such as proto-PNe and PNe
(paper I).  
This indicates that the ring-like features detected in PNe and proto-PNe 
are either not related to the regular mass-loss patterns of AGB stars, or
that the regular AGB patterns are altered or even destroyed at some point
of the subsequent PN formation and evolution.  
To investigate these possibilities, this paper presents a comparative
study of the geometry, surface brightness variations and angular expansion 
of the ring-like structures of three bright, relatively nearby and mature
PNe with high-quality multi-epoch \emph{HST} observations, namely NGC\,6543,
NGC\,7009, and NGC\,7027, with those of the spiral structure around the AGB
star AFGL\,3068.
This study is complemented by radiative-hydrodynamic simulations
to assess the effects of the evolving stellar wind and ionizing
flux during the post-AGB phase on regular patterns produced just
at the end of the AGB phase.
The present paper is organised as follows.
Section~2 describes the four objects and the \emph{HST} observations
used here, Section~3 investigates the geometry of the ring-like
features of these PNe and compares them to the spiral structure of
AFGL\,3068, and Section~4 uses multi-epoch \emph{HST} observations
to investigate the evolution in time of the surface brightness of
these features and their expansion velocity.
Finally, a discussion on the evolution during the post-AGB phase of regular 
patterns resulting from the last gasps of the AGB wind is presented in
Section~5 and some general concluding remarks are laid down in Section~6.

\section{Archival Data and Sample Description} 
\label{sec:obs}

Our previous investigations of ring-like features in proto-PNe and PNe
(paper I) has allowed us to select three PNe, namely NGC\,6543, NGC\,7009,
and NGC\,7027, with multi-epoch \emph{HST} images of sufficient quality to 
detect and trace faint ring-like features around their main nebular shells
and with a time-span adequate to determine their expansion.  
The properties of these three PNe, and these of the carbon star
AFGL\,3068, are presented in Table~\ref{sample}. 
\emph{HST} images of these sources were downloaded from the
Hubble Legacy Archive at the Mikulski Archive
for Space Telescopes (MAST)\footnote{
  \url{http://hla.stsci.edu/}}
to obtain a description of their morphological features
and to investigate their angular expansion.    
Details of the observations used in this paper are listed in 
Table~\ref{obs}.  
According to the keywords  listed in the ``Purpose'' column of this table,
the observations were used to prepare the images presented in this paper
(IMG), to build surface brightness profiles (SB), or to investigate the
angular expansion (EXP).  
To avoid any instrumental effects, the investigation of the angular 
expansion has been carried out using only observations obtained with 
the same \emph{HST} camera and filter.

\begin{figure*}
\begin{center}
\includegraphics[width=\linewidth]{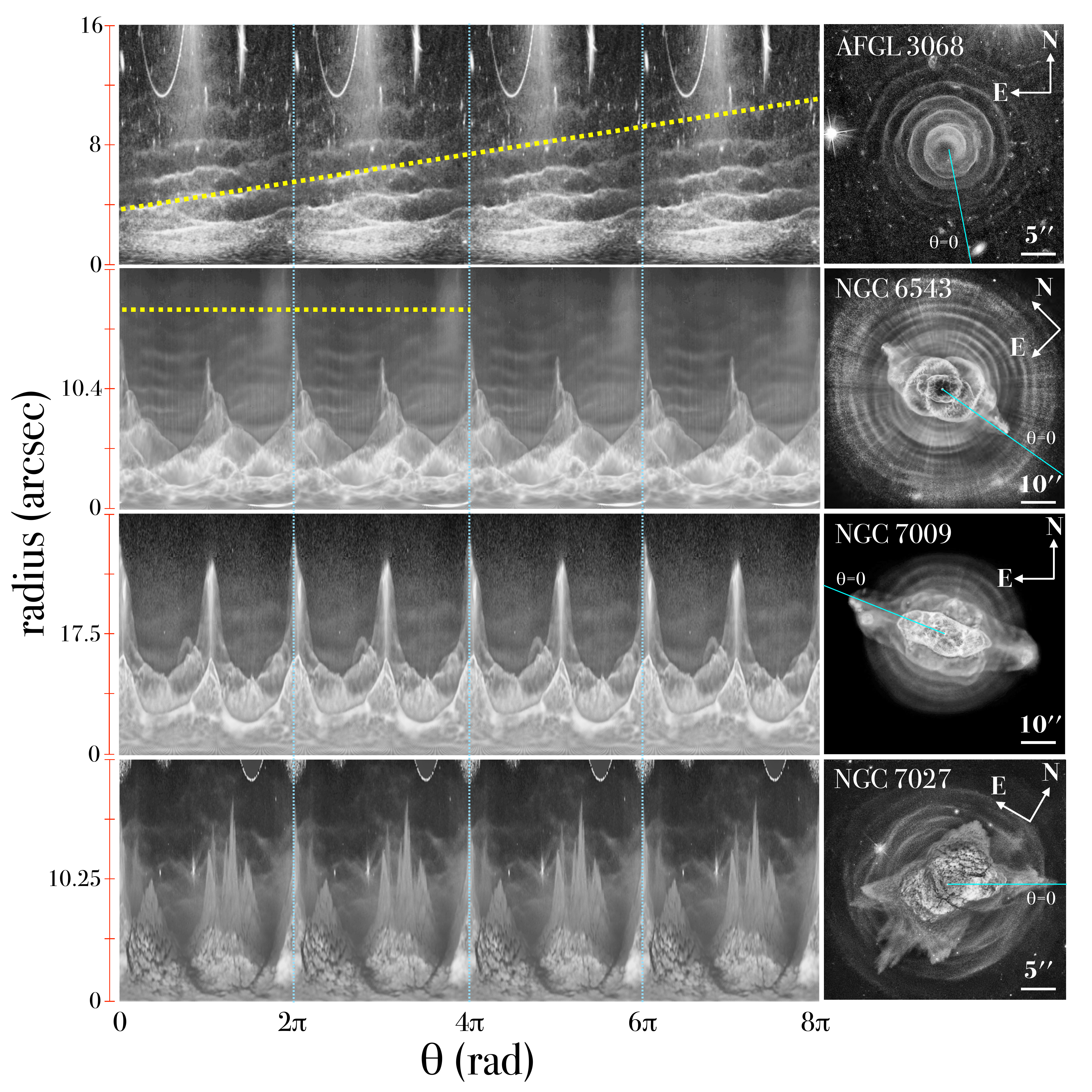}
\caption{
(left)
\emph{HST} images of AFGL\,3068, NGC\,6543, NGC\,7009, and NGC\,7027 presented
in polar ($r - \theta$) coordinate space, where each column has been computed
from a wedge-shaped region with aperture angle of 2$^\circ$ for AFGL\,3068 and
1$^\circ$ for the three PNe.
The vertical extent of these panels is 16\farcs0, 20\farcs8, 35\farcs0,
and 20\farcs5, respectively.
The angular coordinate is replicated 4 times to emphasize the angular trends.  
The vertical dotted lines mark the 2$\pi$, 4$\pi$, and 6$\pi$ lines.  
The dotted yellow lines overlaid on the polar plots of AFGL\,3068
and NGC\,6543 show the pattern expected for an Archimedean spiral
and a circular shell, respectively.
(right)
Direct images of each object, which are grayscale representations of the
images in Figure~\ref{fig:pics}.  
The angular coordinate $\theta$ increases counter-clockwise
and its origin is labeled.
}
\label{fig:polar1}
\end{center}
\end{figure*}

\begin{figure}
\begin{center}
\includegraphics[bb=21 190 584 710,width=0.98\linewidth]{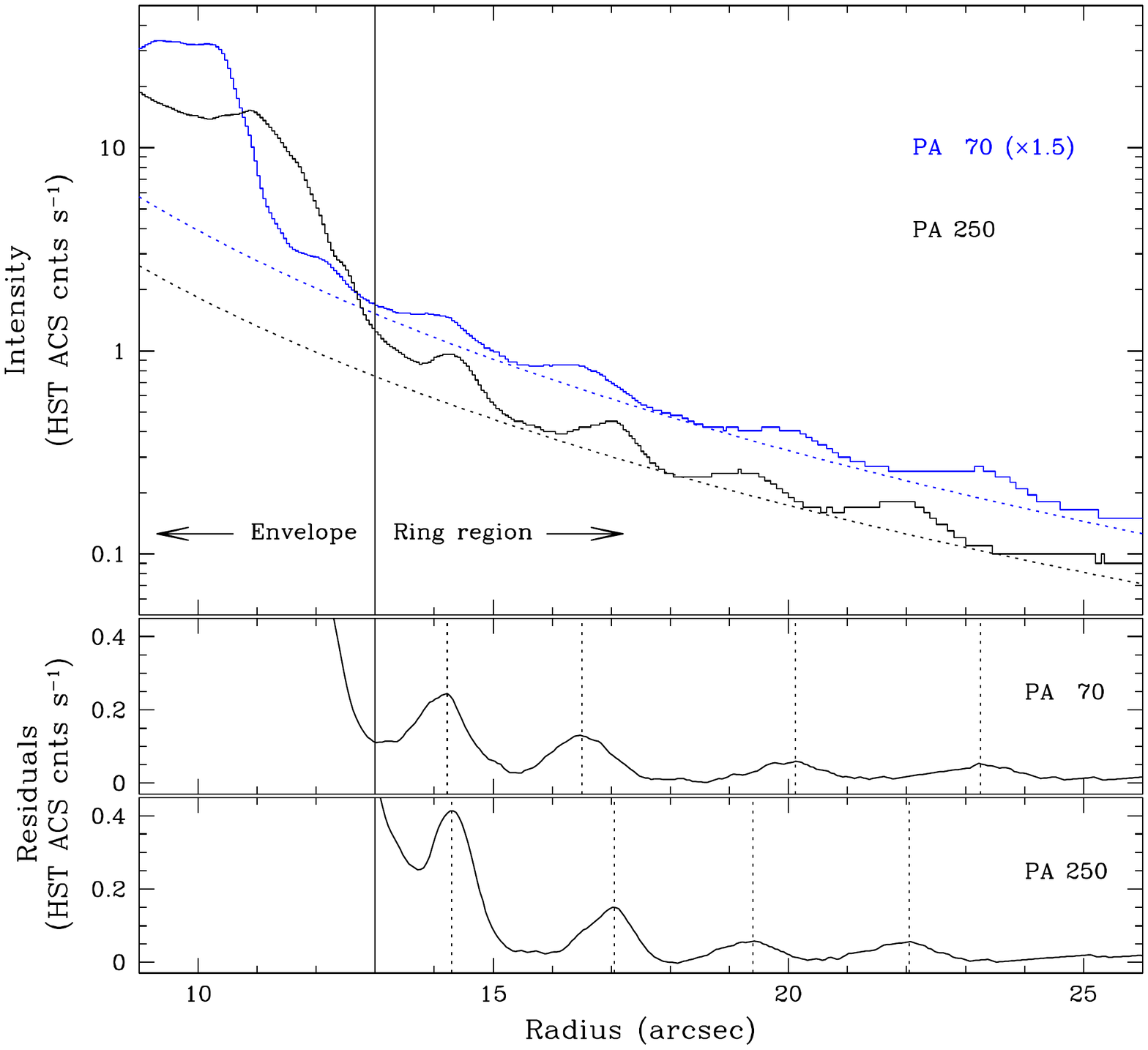}
\caption{[O~{\sc iii}] surface brightness profiles of the Cat's Eye
  Nebula (NGC\,6543) extracted from the \emph{HST} ACS F502N image.
  The top panel shows the radial surface brightness profiles extracted
  along PAs 70$^\circ$ (blue histogram) and 250$^\circ$ (black
  histogram), where the former has been multiplied by 1.5 to make
  easier the comparison of both profiles.  These are obtained using
  wedge-shaped apertures along these PAs with aperture angle of
  10$^\circ$ to improve the signal-to-noise ratio at large radii.  The
  solid vertical line marks the border between the envelope-dominated
  region of the surface brightness profile and the region where rings
  are clearly detected.  The dashed lines show the $r^{-3.6}$ (blue
  line) and $r^{-3.4}$ (black line) fits to these surface brightness
  profiles.  The residuals of these fits are shown in the bottom
  panels for the two PAs.  The vertical dashed lines in these panels
  mark the radial location of each ring.  }
\label{fig:prof}
\end{center}
\end{figure}

Figure~\ref{fig:pics} shows high-quality
colour-composite images of these four objects. Layers, masks and
unsharp masking techniques were applied to the images presented in
this figure to enhance the weak ring-like structures, but we note 
that these images are not used to extract any quantitative  
information besides the spatial distribution of the
ring-like features.  In particular, the image presented here for
NGC\,7009 is a composition of a \emph{HST} WFPC2 picture for the inner
shell and a Very Large Telescope (VLT) MUSE picture for 
the envelope and the ring-like external features.  
The most obvious structural component in
AFGL\,3068 is a spiral pattern, with up to seven turns.  The spatial
distribution of this dust-scattered emission follows an Archimedean
spiral \citep{Mauron2006}, which extends even closer to the central
source in sub-mm CO observations \citep{Kim2017}.
Lacking a fast stellar wind, there is no bright inner shell 
in AFGL\,3068, contrary to NGC\,6543, NGC\,7009 and NGC\,7027, 
whose central stars current fast stellar winds have carved 
inner cavities into their nebular envelopes.

\section{Geometry of the Rings}

The comparison between the shape of the ring-like features around proto-PNe
and PNe and the regular patterns around AGB stars can help us elucidate
whether there is an evolutionary link between these morphological structures.
The pattern around AFGL\,3068 (top-left panel of Fig.~\ref{fig:pics}) 
certainly portrays a spiral, whereas the ring-like features around the
three PNe in the remaining panels in Figure~\ref{fig:pics} cannot be
described neither as circular nor as spiral, in agreement with the 
conclusions presented in Paper I.

These statements are substantiated in Figure~\ref{fig:polar1}, which shows 
the \emph{HST} images of these sources in polar ($r,\theta$) coordinates. 
The polar image of AFGL\,3068 (Fig.~\ref{fig:polar1}-top) clearly shows 
the signature of an Archimedean spiral pattern, for which the radius of 
each emission features increases almost linearly with the angular 
coordinate $\theta$ (the dotted yellow line in Fig.~\ref{fig:polar1}).
This spiral pattern extends from 3\farcs4 up to 17\farcs4
from the position of AFGL\,3068, whereas that seen in ALMA
molecular observations extend from $\simeq$1$^{\prime\prime}$
up to $\simeq$10$^{\prime\prime}$ \citep{Kim2012,Kim2017}.

The innermost regions of the polar images in Figure~\ref{fig:polar1}
of the three PNe are dominated by the emission from their inner rims 
and nebular envelopes. 
The ring-like features are clearly detected 
outside the main nebular shells.  
These show notable wave-like patterns, which are additionally distorted by
collimated protrusions arising from the inner rims and nebular envelopes.
These same wave-like patterns are present in the polar image of
AFGL\,3068, particularly for the outermost regions of the spiral
pattern.  
However, whereas the linear increase of radius with $\theta$ prevails
here, such trend can only be hardly identified in specific regions of
the polar images of the PNe.  
It can be concluded that the ring-like features of these PNe do 
not follow the trend to increase their radius with $\theta$ as in 
Archimedean spirals, yet their shapes also depart from that of 
perfect circles.

Furthermore, the width and spacing of the rings of PNe
change with position angle (PA).
This is revealed in the enhanced images of NGC\,6543, NGC\,7009, and
NGC\,7027 in Figure~\ref{fig:pics}, where the rings towards the West
of NGC\,6543, North-Northwest of NGC\,7009, and East-Northeast of
NGC\,7027 are sharper than those along the opposite directions,
respectively.  This is illustrated into further detail for the case of
NGC\,6543 in Figure~\ref{fig:prof}, which shows surface brightness
radial profiles of the region including rings extracted from the
\emph{HST} ACS F502N [O~{\sc iii}] image along opposite directions 
of particular interest at PA 70$^\circ$ and 250$^\circ$.  
Up to 4 rings can be detected in these spatial
profiles (Fig.~\ref{fig:prof}-top), with the Eastern rings at PA
70$^\circ$ being broader than the Western ones at PA 250$^\circ$,
which are brighter and sharper.  The basal emission from these surface
brightness profile can generally be described by an
$r^{-3.6}-r^{-3.4}$ decline, in close agreement with the
$r^{-3.3\pm0.1}$ fit found by \citet{Balick2001}.  
After subtracting this basal emission (middle and bottom panels in 
Fig.~\ref{fig:prof}), the residuals reveal more clearly the rings.  
The Western rings are indeed sharper and brighter, but these 
plots also disclose that the radial distances of rings located 
along opposite directions are not coincident.  
Measurements of rings distances to the CSPN made on spatial profiles 
extracted along several directions indicate that the inter-ring spacing 
of the Western rings is smaller and more coherent, 2\farcs6$\pm$0\farcs3, 
than that of the the Eastern rings, 3\farcs0$\pm$0\farcs7.
It is pertinent to remark that these asymmetries cannot arise from 
the interaction with the interstellar medium caused by the proper 
motion of the CSPN, as these ring-like features are found inside 
much larger round haloes whose shapes argue against such interactions
\citep{MCW1989,MFG1998,NCM2003}.  
Interestingly, these asymmetries result naturally in eccentric binary 
interactions during the AGB \citep{Kim2019}.

The radial profiles of the outer regions of NGC\,6543 shown in 
Figure~\ref{fig:prof} can also be used to investigate whether 
these rings broaden with their radial distance from the CSPN.  
Gaussians fits to these profiles find that their widths consistently 
increase from the innermost to the outermost ring-like features shown 
in these profiles by $\sim$18\% in the Western side and by $\sim$47\% 
in the Eastern side.
We note that a small fraction of this broadening may be caused
by the average over large spatial regions resulting from the
wedge-shaped aperture used to derive these profiles given the
low signal-to-noise ratio of the emission from these features.  
A detailed analysis of the broadening of rings with radial distance in a
meaningful sample of PNe with deep \emph{HST} observations is mandatory
to strengthen this tantalizing result.

\begin{figure}
\begin{center}
\includegraphics[bb=21 221 592 573,width=0.98\linewidth]{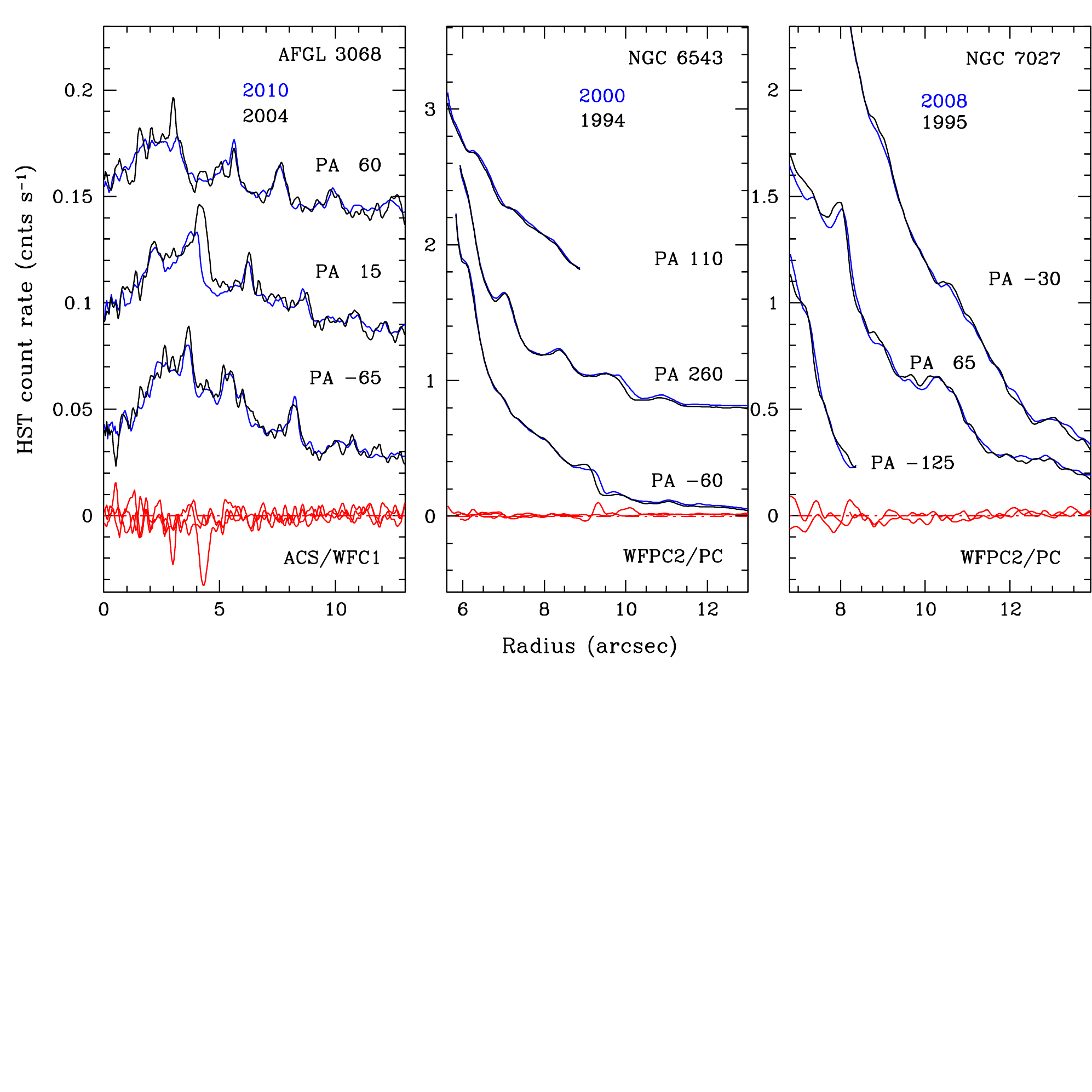}
\caption{
Comparison of multi-epoch surface brightness profiles (blue and black lines) 
of AFGL\,3068, NGC\,6543, and NGC\,7027 along the different radial directions 
labeled on each panel obtained using wedge-shaped apertures with aperture 
angle of 2$^\circ$.  
Profiles along different directions have been shifted vertically to improve 
the readibility of the figure.  
The difference between each profile is shown by the red lines at the bottom 
of each panel.  
}
\label{fig:prof2}
\end{center}
\end{figure}

\section{Time Evolution of the Rings}

Multi-epoch \emph{HST} images can be used to assess the
surface brightness variations and angular expansion of the
ring-like features of the PNe NGC\,6543 and NGC\,7027 and 
the spiral pattern around AFGL\,3068 for which suitable
multi-epoch datasets are available.
NGC\,7009 is thus excluded from the subsequent analysis in this section.  
As noted in \S\ref{sec:obs}, only images obtained with the same \emph{HST} 
camera and filter have been analysed here to avoid instrumental effects.  
The time-lapse between images is $\Delta t=6.09$ yr for AFGL\,3068, 
5.99 yr for NGC\,6543, and 12.98 yr for NGC\,7027.
The possible variations in the surface brightness profiles and the 
general angular expansion caused by the time evolution of the ring-like 
features of PNe and the spiral pattern around AFGL\,3068 are investigated 
in the next two sections.

\subsection{Surface Brightness Variations}

To investigate the surface brightness variations of the outermost structures
of these sources, we have extracted radial profiles along clean directions
from images obtained at two different epochs and compare then in
Figure~\ref{fig:prof2}.  
A preliminary inspection does not reveal any obvious difference in the surface 
brightness radial profiles along these directions.  
These are examined more carefully at the bottom of each panel of 
Figure~\ref{fig:prof2}, where the surface brightness radial profiles of 
different epochs have been subtracted.  
The residuals (shown in red in these figure) are generally flat, indicating 
that no noticeable changes in the surface brightness radial profiles are 
detected in time-scales of a few years.   
Significant variations appear in the innermost regions of NGC\,6543 
and NGC\,7027 (not shown in the figure), but these are attributed to 
the expansion of the inner nebular shell.  
Otherwise, the small fluctuations seen at the location of some 
rings are due to their expansion between different epochs.  
This expansion will be examined into further detail next.

\begin{figure*}
\begin{center}
  \includegraphics[bb=-30 148 630 808,width=0.39\linewidth]{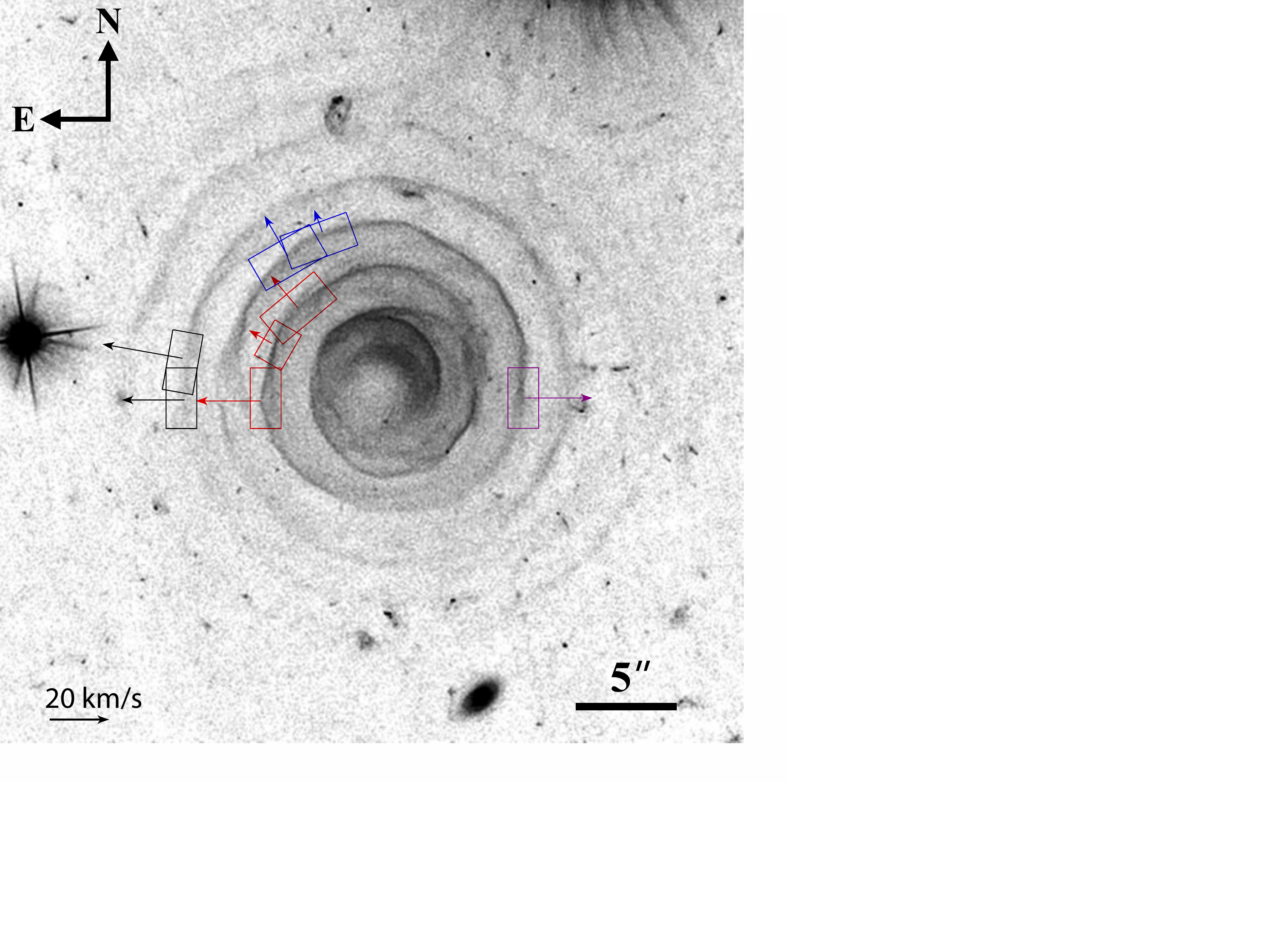}
  \includegraphics[bb=35 244 582 635,width=0.51\linewidth]{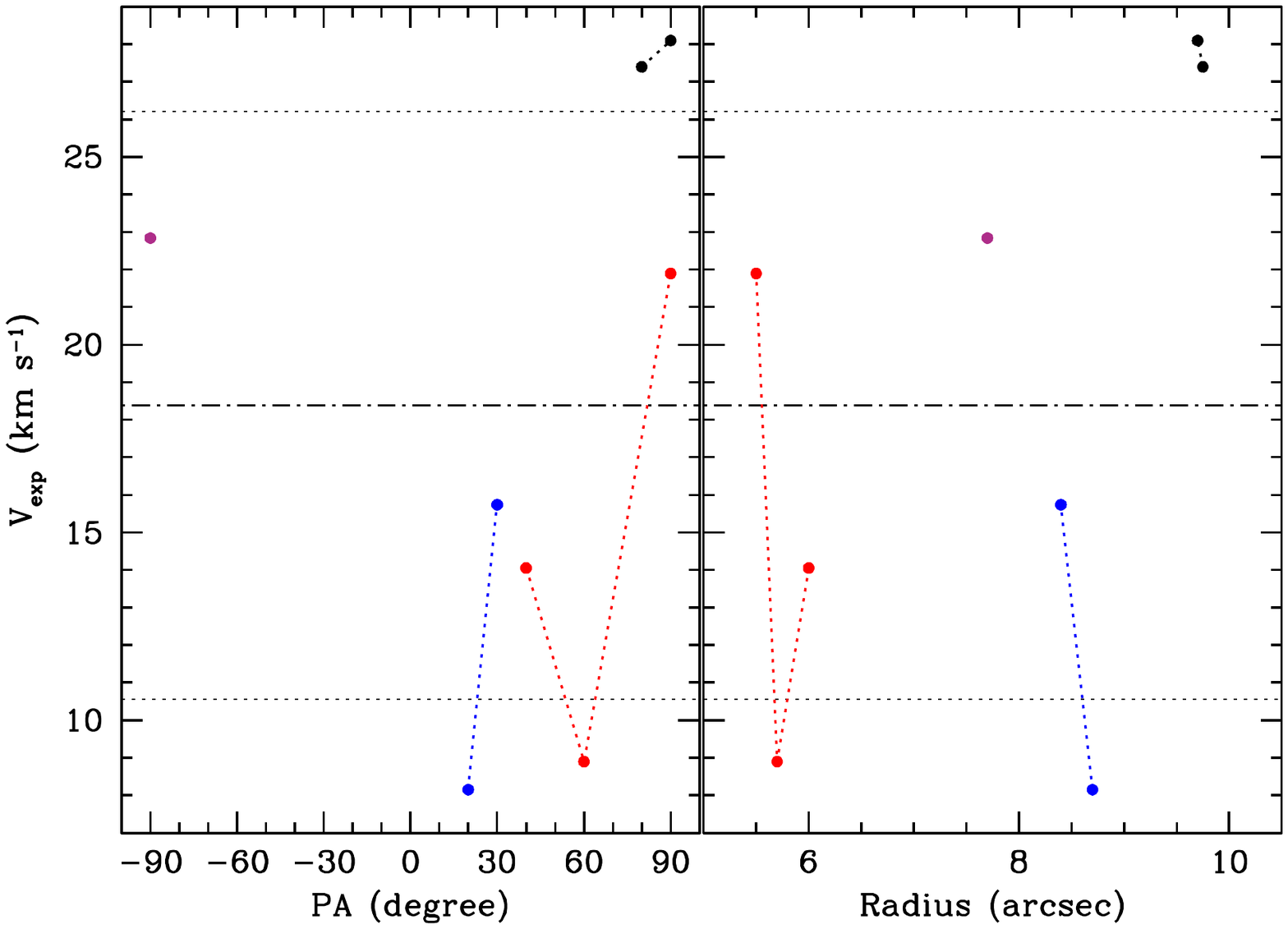} \\
  \vspace*{0.15cm}
  \includegraphics[bb=-30 148 630 808,width=0.39\linewidth]{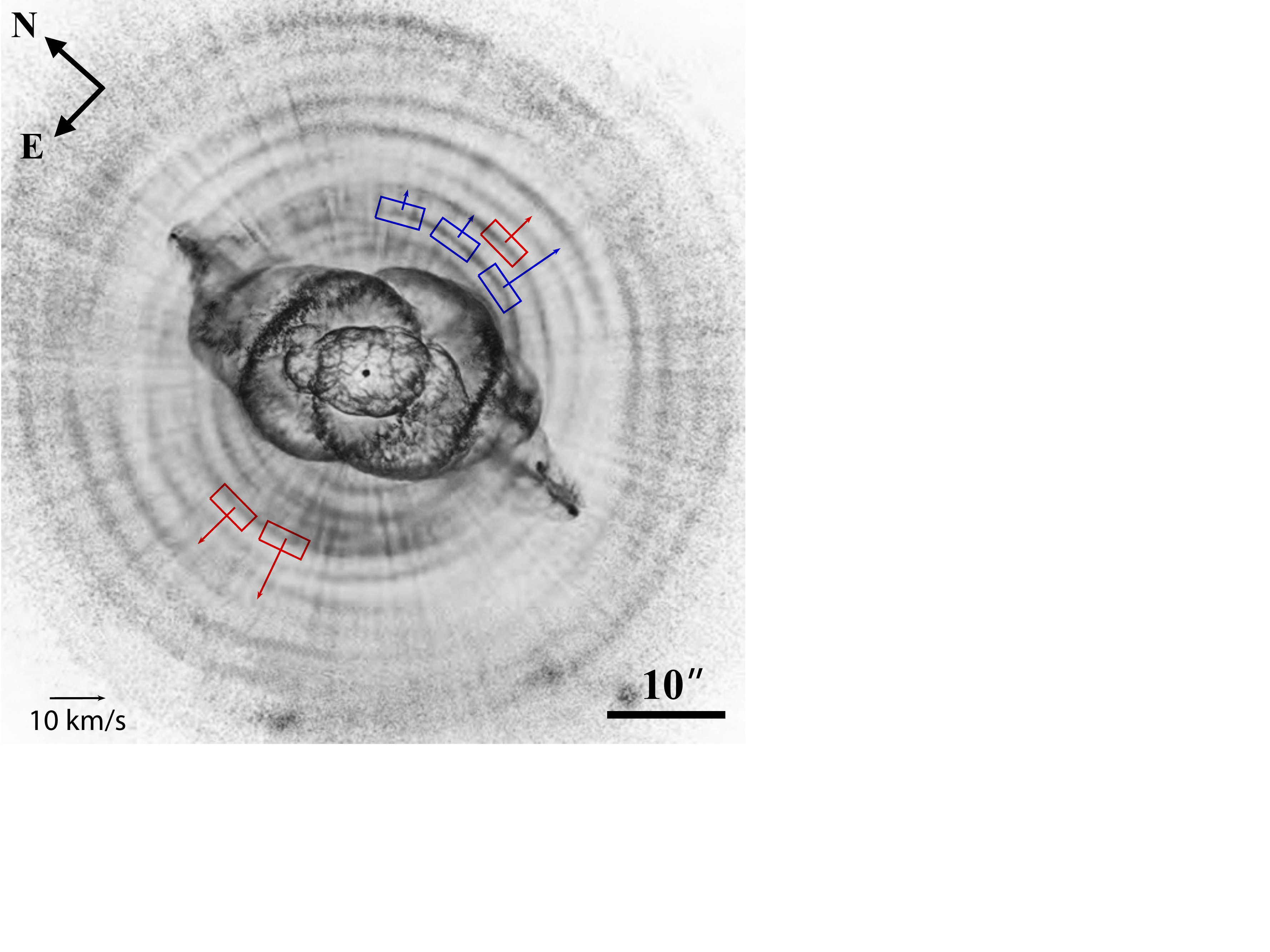}
  \includegraphics[bb=35 244 582 635,width=0.51\linewidth]{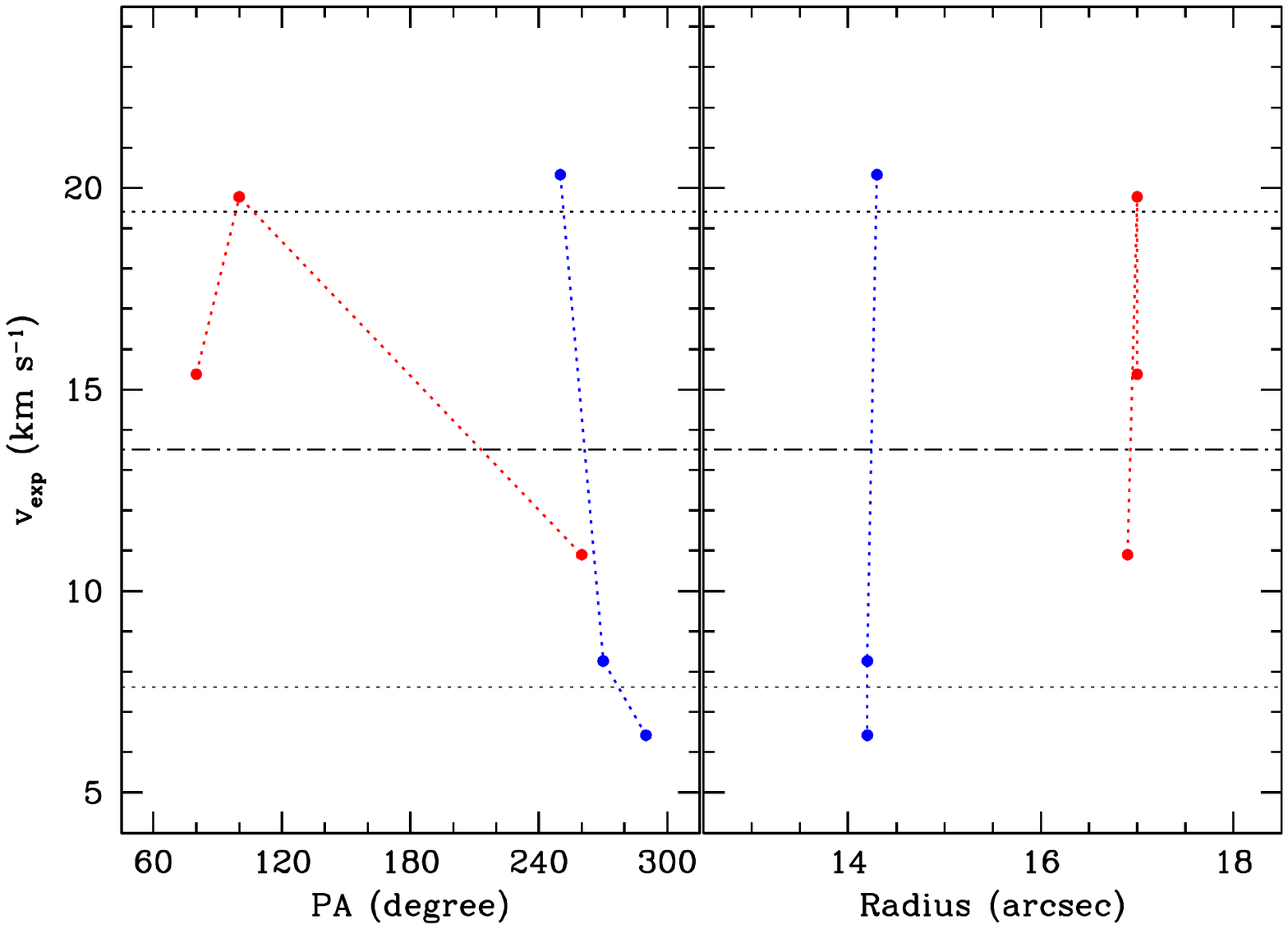} \\
  \vspace*{0.15cm}
  \includegraphics[bb=-30 148 630 808,width=0.39\linewidth]{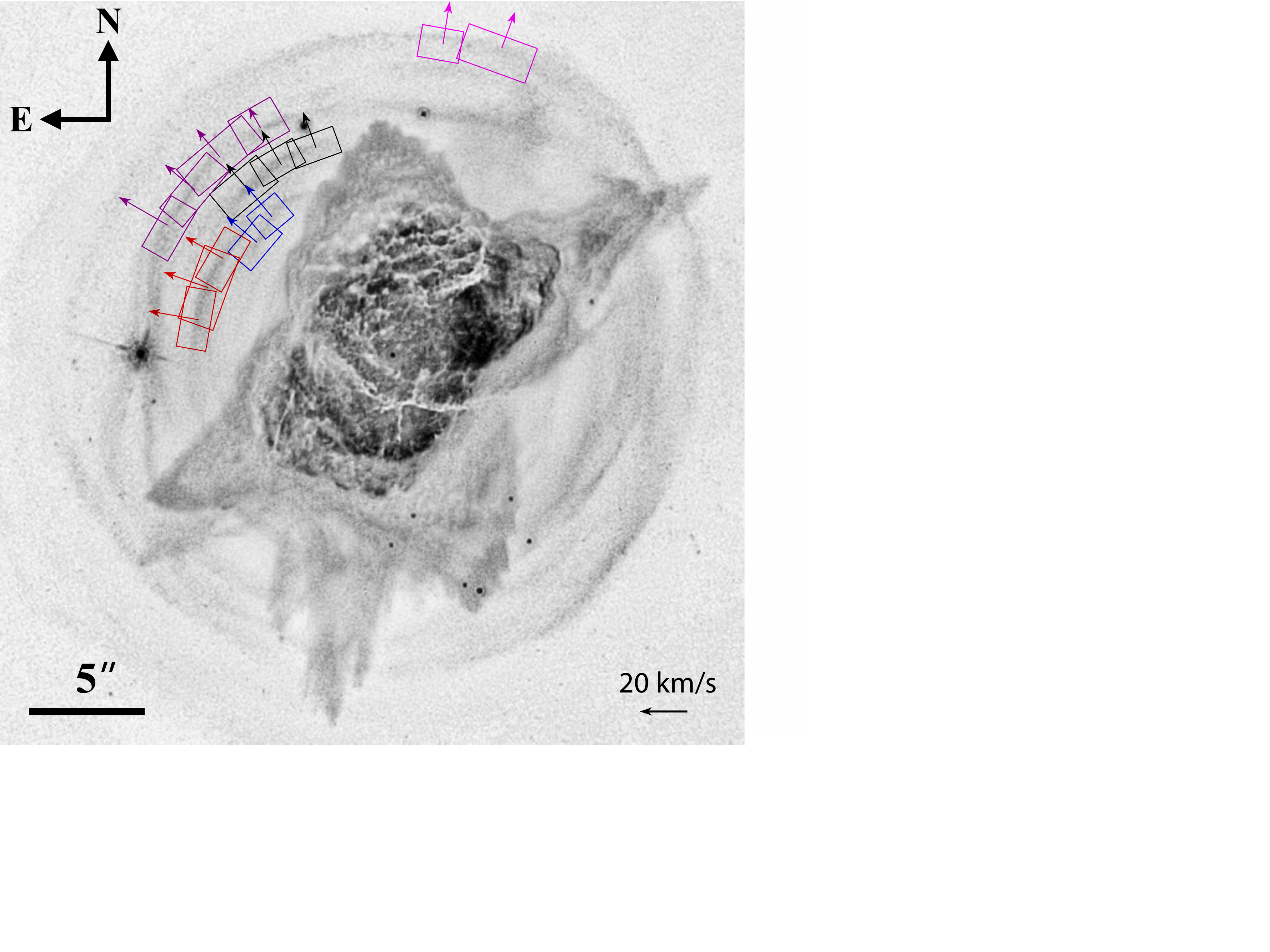}
  \includegraphics[bb=35 244 582 635,width=0.51\linewidth]{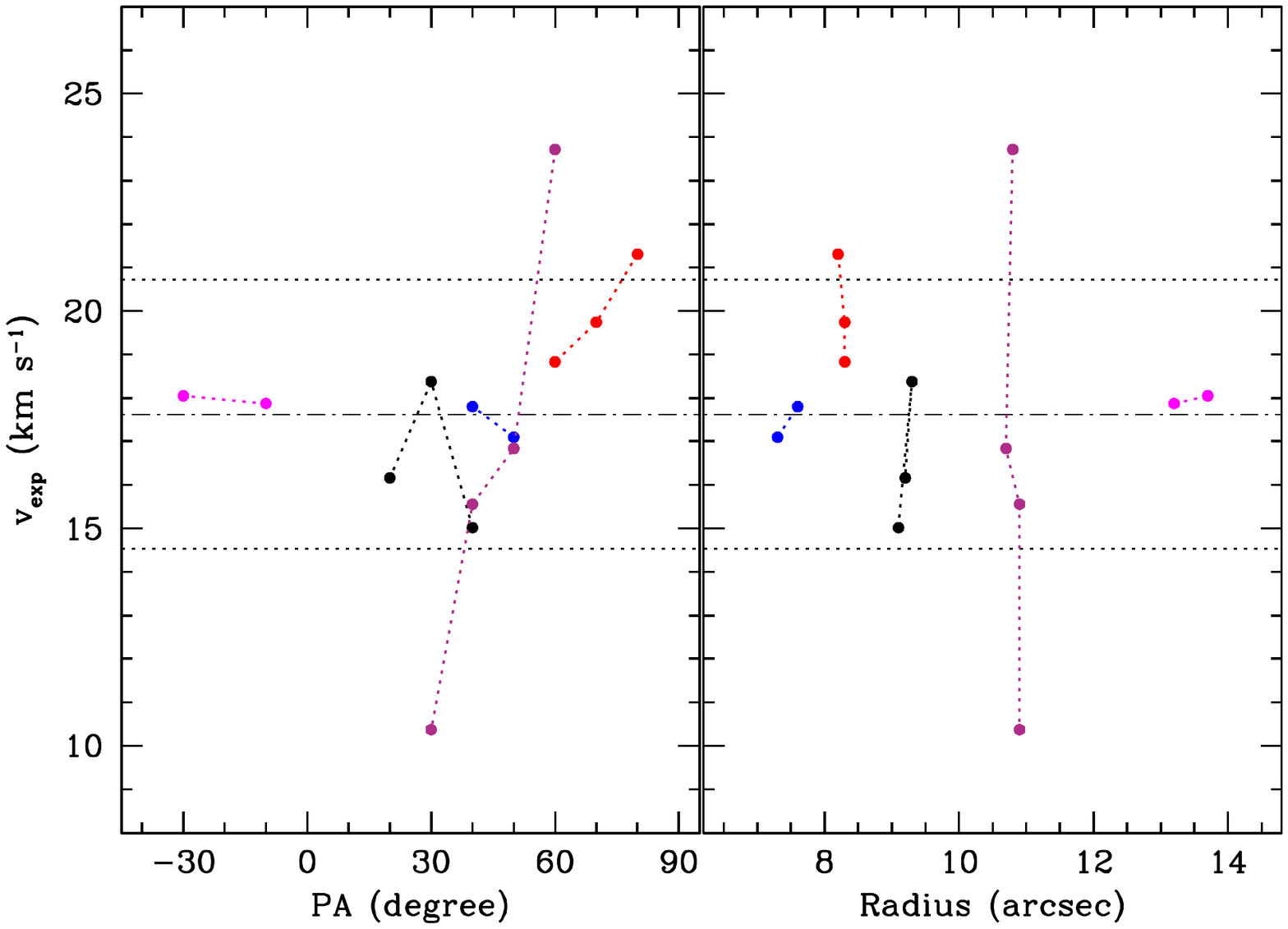}
\caption{
(\emph{left})
Grayscale representations of the images of AFGL\,3068 (\emph{top}),
NGC\,6543 (\emph{middle}), and NGC\,7027 (\emph{bottom}) in
Figure~\ref{fig:pics} showing the rectangular apertures used to derive
the expansion velocities of ring-like features and arms of spiral patterns.
Velocity vectors are represented by arrows with length proportional
to the expansion velocity.
The color of the rectangular boxes on each image denotes 
different arc-like features and spiral arms.
(\emph{right})
Plots showing the dependence of the expansion velocities of
rings and spiral arms of these nebulae with respect
to the PA and radial distance to their central stars.
The color of each data point matches that of the rectangular boxes
enclosing it in the corresponding image in the left panel.
Data points associated to a same arc-like feature or spiral arm
are connected by dotted lines of the same color.  
}
\label{fig.exp}
\end{center}
\end{figure*}

\subsection{Angular Expansion}

The comparison of different epoch high-quality images with stable 
point-spread function (PSF) of PNe, as those obtained by \emph{HST}, 
has been used to confirm and measure the angular expansion rate of
their bright rims and nebular envelopes \citep[see][and references therein
for the most recent use of this method]{SBJ2018}.  
The direct use of this technique to the arc-like features in the haloes
of PNe and spirals around AGB stars is hampered by their extended,
diffuse morphology, their weakness, and the steep surface brightness 
profile of the emission from the PN halo.  

\begin{figure*}
\begin{center}
\includegraphics[bb=10 15 1000 340,width=\linewidth]{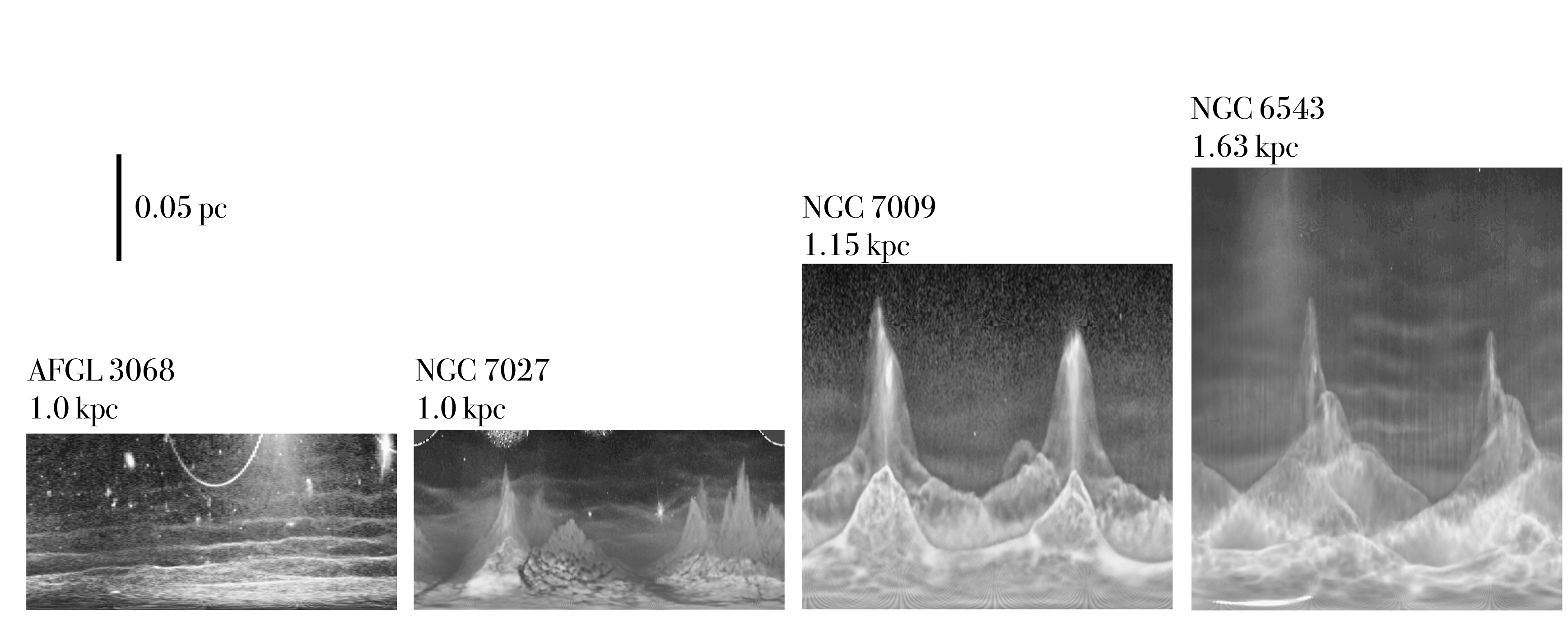}
\caption{
Polar images of AFGL\,3068, NGC\,7027, NGC\,7009, and NGC\,6543.  
This is a revision of Figure~\ref{fig:polar1}, where the radial sizes
have been transformed to physical scale as indicated by the vertical
bar according to the distance to each source.
}
\label{fig:scale}
\end{center}
\end{figure*}

To assess the expansion of these features in AFGL\,3068, NGC\,6543,
and NGC\,7027 we have adopted the magnification method applied by
\citet{SBJ2018} with the refinement introduced by \citet{S_etal2019}
to measure the expansion of the nova shell IPHASX\;J210204.7$+$471015.
Basically, the images from different epochs have been aligned 
using the central stars of NGC\,6543 and NGC\,7027, and galaxies 
in the field of view for AFGL\,3068, and the older image is then
magnified in steps as small as 0.01\%, its flux scaled to that of
the newest image, and subtracted from the latter.
The count number standard deviation of the pixels inside the rectangular
apertures shown in the left panels of Figure~\ref{fig.exp} are then computed.
These apertures have typical sizes of 1\farcs0--1\farcs5 by
$4^{\prime\prime}$--$7^{\prime\prime}$ and are oriented along the
arc-like features of NGC\,6543 and NGC\,7027 and the spiral
arms of the pattern around AFGL\,3068.
They trace the sharpest features registered in the
images obtained at different epochs, avoiding field 
stars and hot pixels or any other image blemish.
We note that some apparent sharp features are unfortunately affected
by image imperfections or have not been registered by the observations
obtained at the two different epocs.
These have been disregarded from further calculations.
The standard deviation of the pixels inside these rectangular apertures,
when plotted against the angular offset of the oldest image (i.e., the
magnification factor times the radial distance of the rectangular aperture
to the CSPN) reveals typical patterns
that decrease, reach a minimum value, and then increase (Appendix~A).  
The minimum value of these curves has been used to determine
the angular offset that best fit the angular expansion of the
feature inside each rectangular aperture.

The angular expansion rate of the ring-like features of NGC\,6543 and 
NGC\,7027 and the spiral pattern of AFGL\,3068 at the location of these 
apertures can be computed from this best fit angular offset and the
time-lapse between the different epoch images.  
Adopting the distances to each source listed in Table~\ref{sample},
the expansion velocities at these locations can be derived.
These are shown by arrows in the left panels of Figure~\ref{fig.exp}.
The possible dependence of the expansion velocity with PA
or radial distance is investigated in the right panels of
Figure~\ref{fig.exp}, but results are inconclusive: all 
measurements are within 1-$\sigma$ deviation (the dotted
lines in these plots) of the averaged values of the expansion
velocity on the plane of the sky (the dotted-dashed lines).  
These are
18$\pm$8~km~s$^{-1}$ for AFGL\,3068,
13.5$\pm$6~km~s$^{-1}$ for NGC\,6543, and 
17.6$\pm$3.1~km~s$^{-1}$ for NGC\,7027.  
The expansion velocities derived for these features imply relatively
coherent time-lapses between concentric features:
 530$\pm$150 yr for AFGL\,3068,
1900$\pm$1100 yr for NGC\,6543, and 
 520$\pm$140 yr for NGC\,7027.  
Previous estimates of these for AFGL\,3068 and NGC\,6543
yield 800 and 1500 yr, respectively \citep{Kim2017,Balick2001}.

\section{From Regular Patterns Around AGB Stars to Arcs Around PNe}

The prevalence of incomplete rings and arcs outside the nebular core 
of PNe was evidenced in Paper~I.  
On the contrary, the occurrence of spiral patterns
\citep{Kim2012,Decin2015,Kim2017,Homan2018} and
regular shells \citep{GD2001,Mauron2006} among AGB
stars is quite significant.  
The analysis of the \emph{HST} images of AFGL\,3068, NGC\,6543, NGC\,7009,
and NGC\,7027 presented in previous sections provides us with interesting
hints to understand the connection and possible evolution between the regular
patterns surrounding AGB stars and irregular arcs around PNe.

The expansion velocity on the plane of the sky of the spiral pattern around
AFGL\,3068 derived from multi-epoch \emph{HST} images of dust-scattered 
light in the previous section (18$\pm$8 km~s$^{-1}$) is consistent with the 
radial velocity expansion derived from ALMA data of CO molecular emission 
\citep[14 km~s$^{-1}$;][]{Kim2017}, indicating that the same expansion 
pattern is detected by these independent methods.  
This expansion velocity is similar to that measured in
the spiral patterns and shells around AGB stars such as
EP\,Aqu \citep[14 km~s$^{-1}$;][]{Homan2018},
IRC+10216 \citep[14.5 km~s$^{-1}$;][]{Cernicharo2015},
RW\,LMi \citep[16 km~s$^{-1}$;][]{Kim2015}, and
R\,Scu \citep[15 km~s$^{-1}$;][]{Maercker2012}, 
suggesting that this is a typical expansion velocity for
material expelled during this phase of stellar evolution
\citep{RSO2009,IE2010}.
The expansion velocities of the arc-like features around NGC\,6543
(13.5$\pm$6~km~s$^{-1}$) and NGC\,7027 (17.6$\pm$3.1 km~s$^{-1}$)
are within a similar velocity range.
Certainly the statistics is very poor, but there is no
reason for the time being to reject the hypothesis that 
the incomplete features seen in the outer shells of PNe 
result from the evolution of regular patterns around AGB 
stars.

The similarity between the time-lapse derived for concentric features
in NGC\,6543 and NGC\,7027 and those derived in AGB stars reinforces
this hypothesis.
The time-lapse between concentric features in NGC\,6543 and NGC\,7027
are 1400$\pm$1100~yr and 520$\pm$140 yr, respectively, whereas in
AGB stars these are derived to be 530$\pm$150 yr (this work) and 800
yr \citep{Kim2017} in AFGL\,3068, 800--1000 yr in IRC+10216
\citep{Cernicharo2015}, and 350 yr in R\,Scu \citep{Maercker2012},
but notably smaller $\simeq$50 yr in EP\,Aqu \citep{Homan2018}.
This comparison is statitically strenghtened with the large sample
of proto-PNe and PNe analyzed in Paper~I, that revealed time-lapses
for these sources in the range from 90 to 2000 yr.

If the regular patterns present in the stellar wind of AGB stars
evolve into the irregular, incomplete concentric patterns detected
in PNe, these morphological changes should arise as the result of
their expansion and of their interaction with the post-AGB stellar
winds and ionization or excitation by the increased UV photon flux
from the CSPN.
These are investigated in the next sections.

\begin{figure*}
\begin{center}
\includegraphics[bb=1 120 1024 660,width=\linewidth]{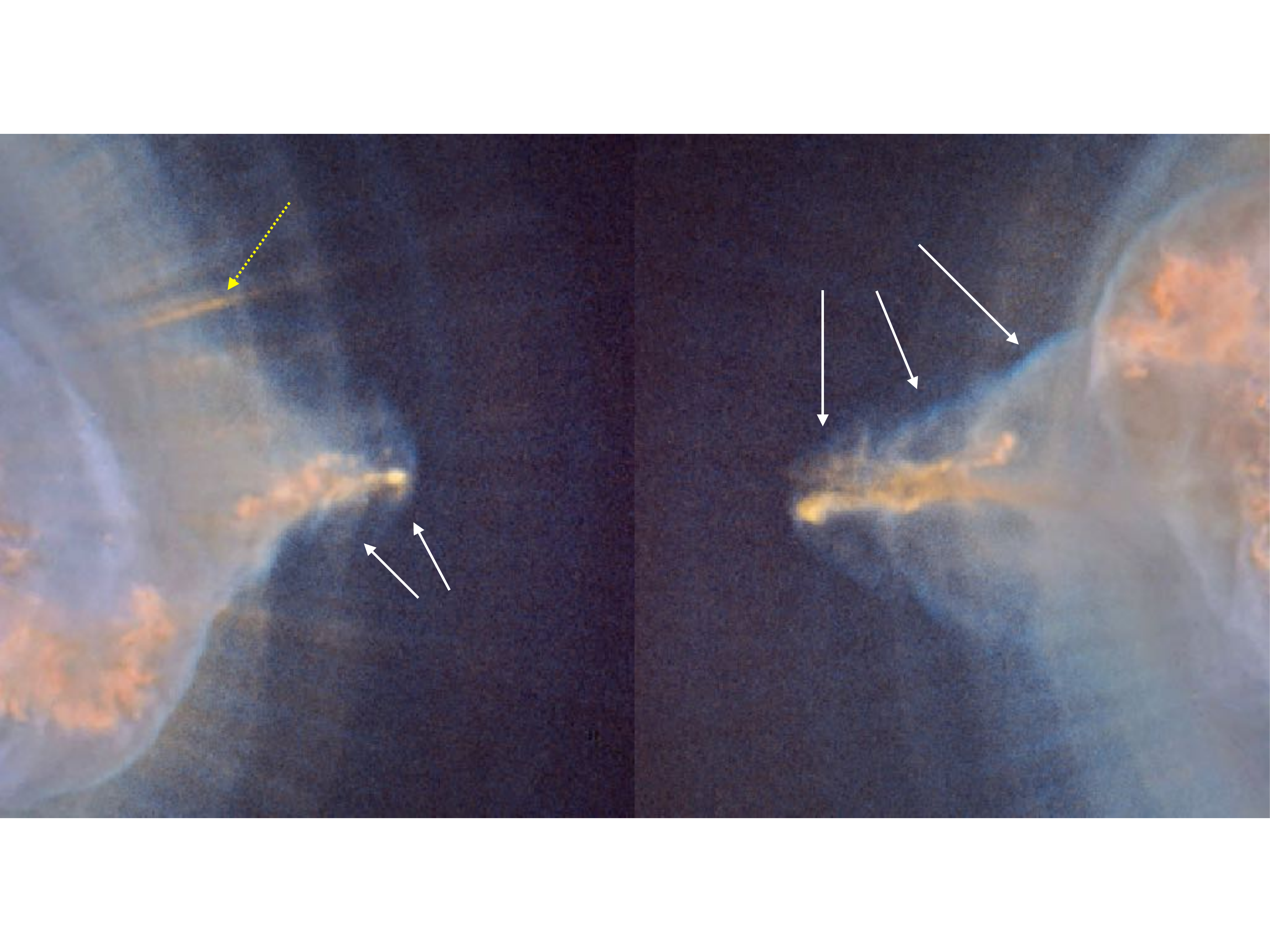}
\caption{
Colour-composite pictures in H$\alpha$ (green), [O~{\sc iii}] (blue)
and [N~{\sc ii}] (red) of the interaction regions between the bipolar
jets of NGC\,6543 and the ring-like structures surrounding the nebular
shell.
White arrows point to bow-shock-like features resulting from the
progression of the bipolar jets through the ring-like structures,
whereas the yellow arrow marks a radial structure of lower ionization
degree across these ring-like features.
}
\label{fig:ngc6543}
\end{center}
\end{figure*}

\subsection{Size Matters}

We present in Figure~\ref{fig:scale} a comparison of the physical size
of AFGL\,3068, NGC\,6543, NGC\,7009, and NGC\,7027 using images in polar
coordinates.  
The different panels, corresponding to different sources, reveal 
that the outermost structures detected in AFGL\,3068 have similar 
physical sizes to the innermost of these structures detected in 
PNe, whereas the outermost features of the most evolved sources, 
NGC\,6543 and NGC\,7009, have sizes about twice larger than those 
of AFGL\,3068.

The farthest regions of the spiral pattern of AFGL\,3068 traced by
the \emph{HST} image show noticeable kinks with amplitude increasing
with radial distance to the central star.  
Actually, the outermost features of AFGL\,3068 look more similar to
the arcs and ring-like features of these three PNe than to the almost 
perfect Archimedean spiral in ALMA observations \citep{Kim2015} and the
innermost regions of the spiral pattern in the \emph{HST} image.
At least for AFGL\,3068, there is a trend for the regular AGB
pattern to get distorted in shape as its distance from the central 
star (and thus its age) increases. 
This can be expected as initial velocity fluctuations gets
amplified in time with the expansion of the spiral pattern
\citep{Kim2019}.

Adopting the average expansion velocity derived from their angular
expansion in \S4.2, the age of the features closest to the central
stars is 
$\sim$700 yr for AFGL\,3068, 
$\sim$6,000 yr for NGC\,6543, and 
$\sim$2,000 yr for NGC\,7027.
If these features expanded thermally, they should thicken as they age, thus 
these relatively large ages, particularly for NGC\,6543 (but also for other 
sources in Paper~I, with ages up to 4500 yr), would imply a noticeable 
broadening of the rings.  
Indeed, the width of the ``younger'' features around AFGL\,3068 are sharper 
than the ``older'' features around NGC\,6543 in Figure~\ref{fig:scale}.
Considering the distance to both sources, the spiral arms around
AFGL\,3068 have a width ($\sigma$) of (3.6$\pm$1.1)$\times$10$^{15}$
cm, whereas the Western arc-like features around NGC\,6543 have a
width of 1.5$\times$10$^{16}$ cm and the Eastern ones have a width
of 1.8$\times$10$^{16}$, i.e., 4--5 times broader than those of 
AFGL\,3068.  
There is no clear variation of the width of the spiral arms 
around AFGL\,3068 with their radial distance to the CSPN, but 
there is a tantalizing indication that the ring-like features 
around NGC\,6543 thicken as they propagate outwards.

\subsection{Breaking Bad}

The inner regions of the polar images of NGC\,6543, NGC\,7009, and NGC\,7027
presented in Figures~\ref{fig:polar1} and \ref{fig:scale} also reveal the
progression of the inner rims, nebular envelopes and collimated outflows
throughout the outer ring-like features.
The outer edge of the nebular envelope drives a shock into the
AGB wind, which is detected through its enhanced [O~{\sc iii}]
to H$\alpha$ ratio \citep{Getal_2013}.  
Any feature in the AGB wind is swept up and destroyed wherever
the nebular envelope reaches them.

Similarly, the expansion of collimated outflows has deep dynamical and 
morphological effects on the ring-like features as shown in
Figure~\ref{fig:ngc6543} 
for the case of NGC\,6543.
The bipolar jets progress through the ring-like structures surrounding 
the nebular shell and produce a series of bow-shock-like features
(the most noticeable ones are marked with white arrows in
Fig.~\ref{fig:ngc6543}), which can be associated with each of these
ring-like structures.  
Similar features are detected in the image of NGC\,7009.
Apparently, the bipolar jets break through a region of
enhanced density every time they cross the boundary of
a ring-like features.
As the ambient gas is deflected from the collimated outflow tips,
it becomes hot and its increased thermal pressure produces a
bow-shock-like feature at the leading tip, while the gas left in
the wake cools adiabatically as it slows down \citep{Betal2013}.  
Since the bipolar jets in NGC\,6543 are not contained on the plane 
of the sky \citep{MS92}, this confirms that the arcs and incomplete 
rings around NGC\,6543 are the projection of 3D structures.
Even if these are 3D structures and this interaction takes place
only at a given inclination angle, their effects in the morphology 
are quite noticeable.

\subsection{Testing the Survival of Regular AGB Patterns}

\begin{figure*}
\begin{center}
  \includegraphics[width=0.8\linewidth]{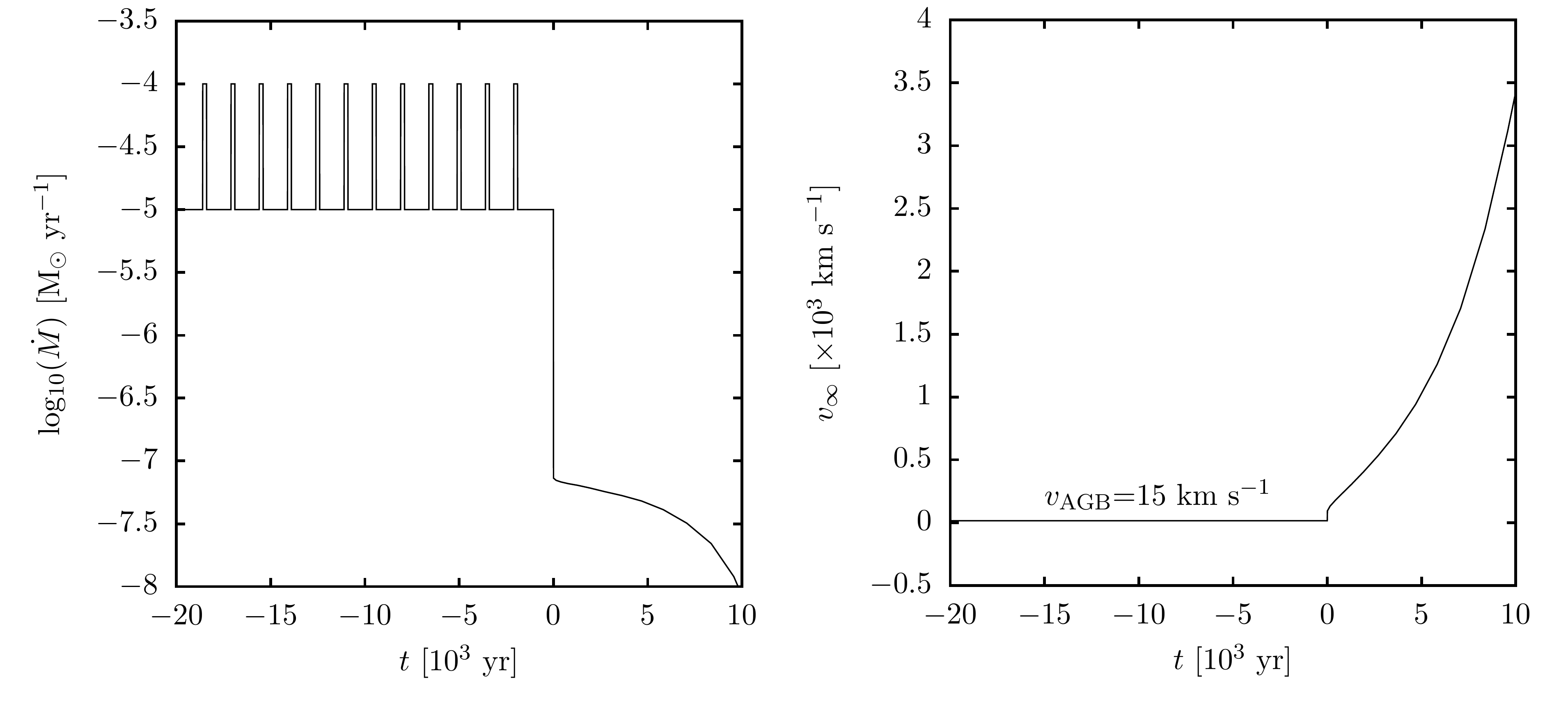}
  \vspace*{-0.5cm}
  \caption{
 Mass-loss rate $\dot{M}$ (left) and stellar wind velocity $v_{\infty}$
(right) as a function of time from the onset of the post-AGB wind.
  }
\label{fig:sim}
\end{center}
\end{figure*}

There is a large number of high-quality 3D numerical works addressing 
the formation and evolution of regular structures and very specifically 
spiral patterns by common envelope binary interactions or eccentric binary
interactions during the AGB phase 
\citep[see][and references therein]{Chamandy2018,Iaconi2018,Kim2019}. 
However, none of these simulations explore the effects on these 
structures caused by the increase of the UV flux from the central star 
and the mechanical luminosity and linear momentum of its stellar 
wind during the post-AGB phase.  
These effects were explored using 1D simulations by \citet{Meijerink2003}, 
who found that the ionizing photon flux from the CSPN is capable to erase 
the signature of regular structures within a few thousand years.

The alternating rays of ionized material intersecting the ring-like 
features and arcs of NGC\,6543 and NGC\,7009 (Fig.~\ref{fig:pics} 
and yellow arrow in Fig.~\ref{fig:ngc6543}) reveal that the effects 
of ionizing photon flux from the CSPN are far from isotropic.  
These rays of ionization arise from shadow effects 
\citep[see][]{Arthur2006,Williams1999} caused by 
clumps and filaments produced as the result of 
Rayleigh-Taylor and thin-shell instabilities in 
the wind-wind interaction region 
\citep[e.g.,][]{Stute2006,Toala2016} as these 
clumps trap the ionizing UV flux from the CSPN 
preventing the ionization to be uniform.
These are effects that can be noticeable even before the onset of ionization,
as illustrated by the fragmented arc segments in the proto-PN CRL\,2688  
\citep{Balick2012}.

These are effects that cannot be modeled using 1D simulations
\citep[as those presented by][]{Meijerink2003} and are specifically
investigated here via radiation-hydrodynamic simulations using the
code described in \citet{Toala2011} and \citet{Arthur2012}.
Here we will adopt a simplified mass-loss rate history for the
last 10$^5$ yr in the AGB phase with a basal mass-loss rate of
$\dot{M}$=10$^{-5}$~M$_{\odot}$ yr$^{-1}$ and a constant wind
velocity of $v_\mathrm{AGB}$=15~km~s$^{-1}$ suitable for a 1.5
M$_\odot$ progenitor \citep{VW93,VW94,Blocker1995}.
In the last 20,000 yr of evolution, the mass-loss rate is artificially 
enhanced by an order of magnitude during 200 yr every 1500 yr to create 
density enhancements between the rings and inter-ring regions.  
The duration and period have been adopted to fit the surface 
brightness variations observed in the ring-like features around
NGC\,6543 (Fig.~\ref{fig:prof}).  
The density enhancement of an order of magnitude has been adopted to match 
the apparent "mass-loss fluctuations" from a few up to several hundreds, 
with time-lapses of a few hundred years due to quasi-periodic mass loss
caused by dust drift on the otherwise constant AGB mass-loss \citep{Simis2001}.
Density enhancement from a few up to one hundred in exceptional cases 
are also found by \citet{Kim2019} for the spiral and inter-spiral
regions in their simulations caused by the eccentric-binary interactions
during the AGB.
Observational determinations of the density contrasts are also in
the range 3--10, as reported for the regular structures surrounding
the AGB stars CW\,Leo and IRC~+10\,218 \citep{Decin2015,Guelin2018}.

The mass-loss rate history and the evolution of the stellar wind velocity
with time used in our simulation are shown in Figure~\ref{fig:sim}.
These are used to produce 1D calculations with 2000 uniformly spaced radial
cells that correspond to 0.5 pc in spatial size, i.e., 2.5$\times$10$^{−4}$~pc
per cell, where the wind injection zone comprises the innermost 10~cells.
No photoionization is considered during this phase.
The final density profile at the end of the AGB phase exhibits the
typical $\rho \propto r^{-2}$ distribution \citep[see, e.g.,][and
references therein]{Villaver2002,Perinotto2004,Toala2014} with
density enhancements overimposed.
Although mass loss pulses are assumed to be periodic, the
separation between the density enhancements in our simulations
is rather quasi-periodic.
This occurs because the material ejected during the late AGB phase,
and particularly that corresponding to each pulse of enhanced mass
loss, interacts hydrodynamically with the material previously ejected.
As a result, the distance between each pulse of enhanced mass
loss is reduced and their separation is not strictly regular.

To study the hydrodynamical effects at the wind-wind interaction zone
and ionizing flux in the post-AGB phase, we remapped the 1D distributions
of density, momentum and total energy at the end of the AGB phase into 2D
cylindrically symmetric grids.
Using these as initial conditions, we obtained 2D simulations 
injecting an evolving post-AGB wind and ionizing photon flux 
from an 1.5 M$_{\odot}$ progenitor star model \citep{Toala2014}.
The 2D axisymmetric numerical simulations are performed on a fixed
grid of 1200 radial by 2400 $z$-direction cells of uniform cell size
and total grid spatial size of 0.3$\times$0.6~pc$^{2}$, i.e., the
1D and 2D simulations have the same radial resolution.
The free-wind injection zone has a radius of 40~cells, which
corresponds to the innermost 0.01~pc.
The 2D simulations start at $t=0$ with the injection of the
post-AGB (fast) wind (Figure~\ref{fig:sim}).
Our numerical results will focus on the innermost 0.3~pc of the
PN where the rings are distributed (Fig.~\ref{fig:scale}).  
This means that we do not follow the evolution of the ionized nebular
shell as it leaves the computational domain.
Note also that, even though the calculations are performed in a full
$r - z$ plane, we only show half of the computational domain.

\begin{figure*}
\begin{center}
  \includegraphics[width=0.33\linewidth]{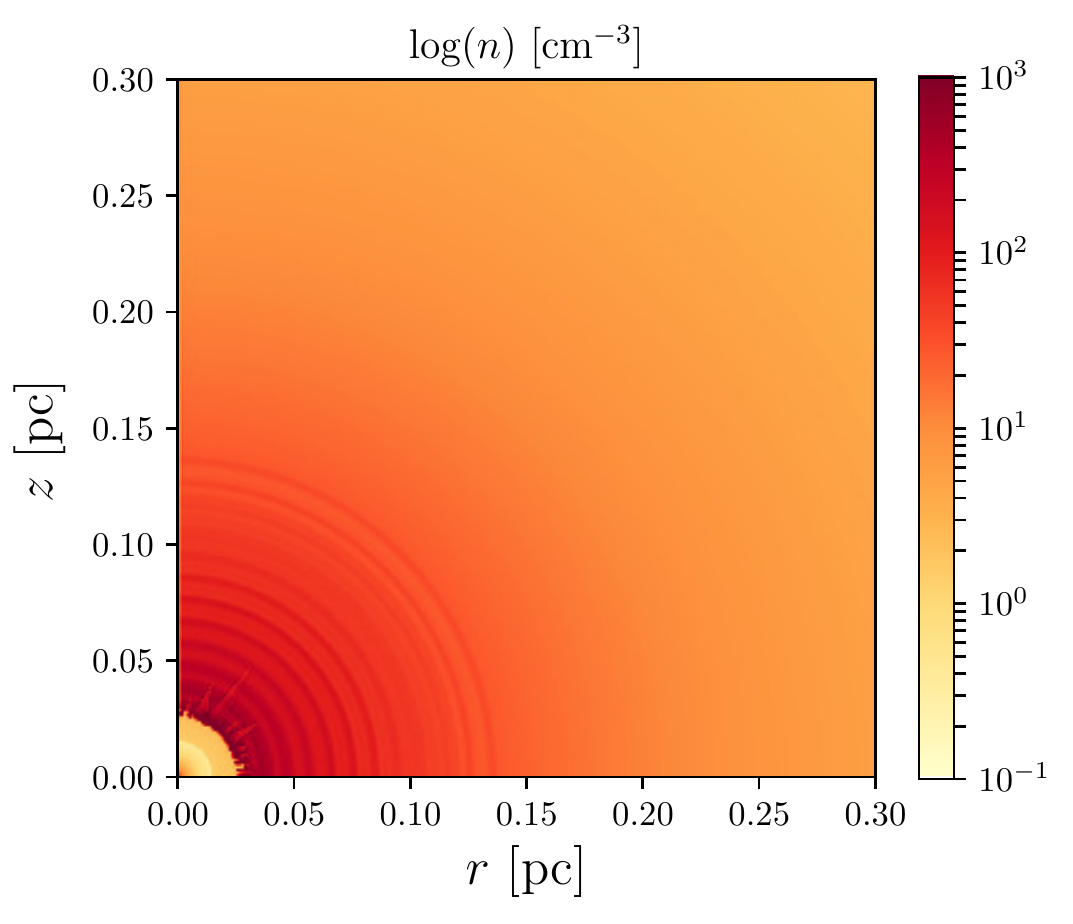}~
  \includegraphics[width=0.33\linewidth]{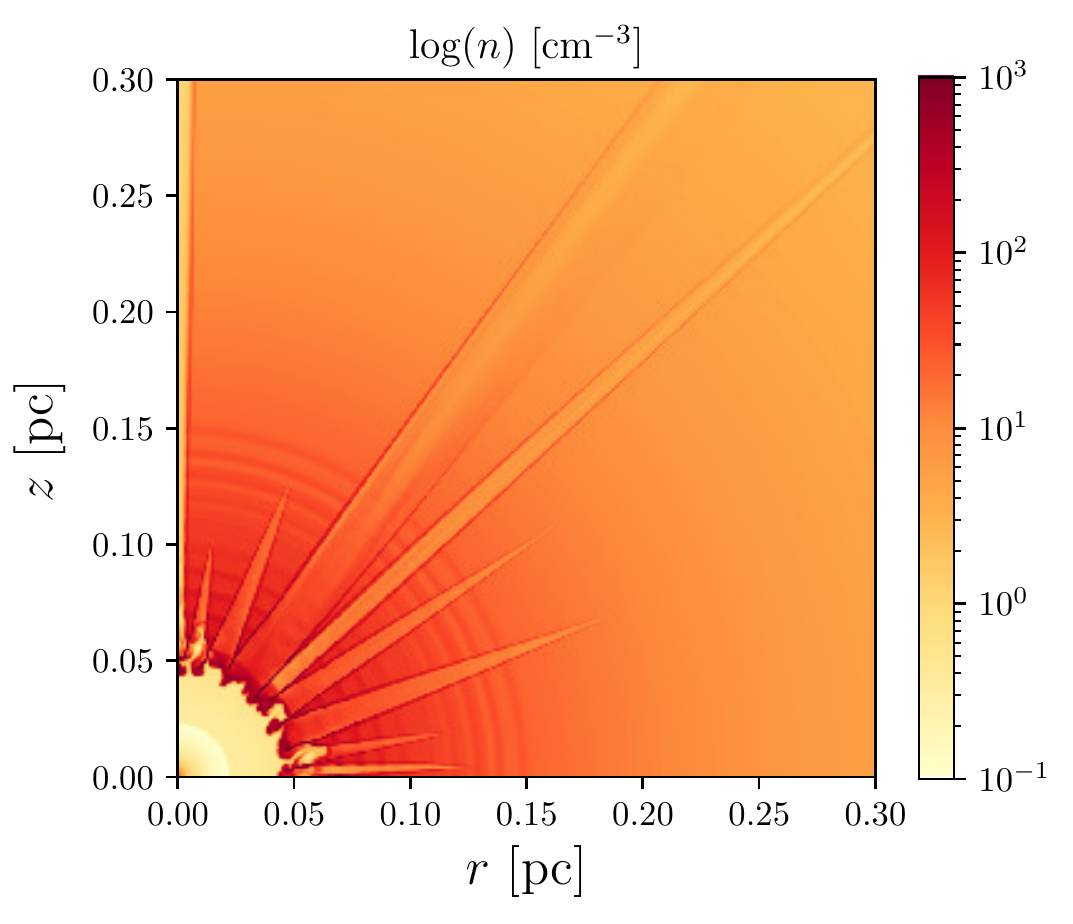}~
  \includegraphics[width=0.33\linewidth]{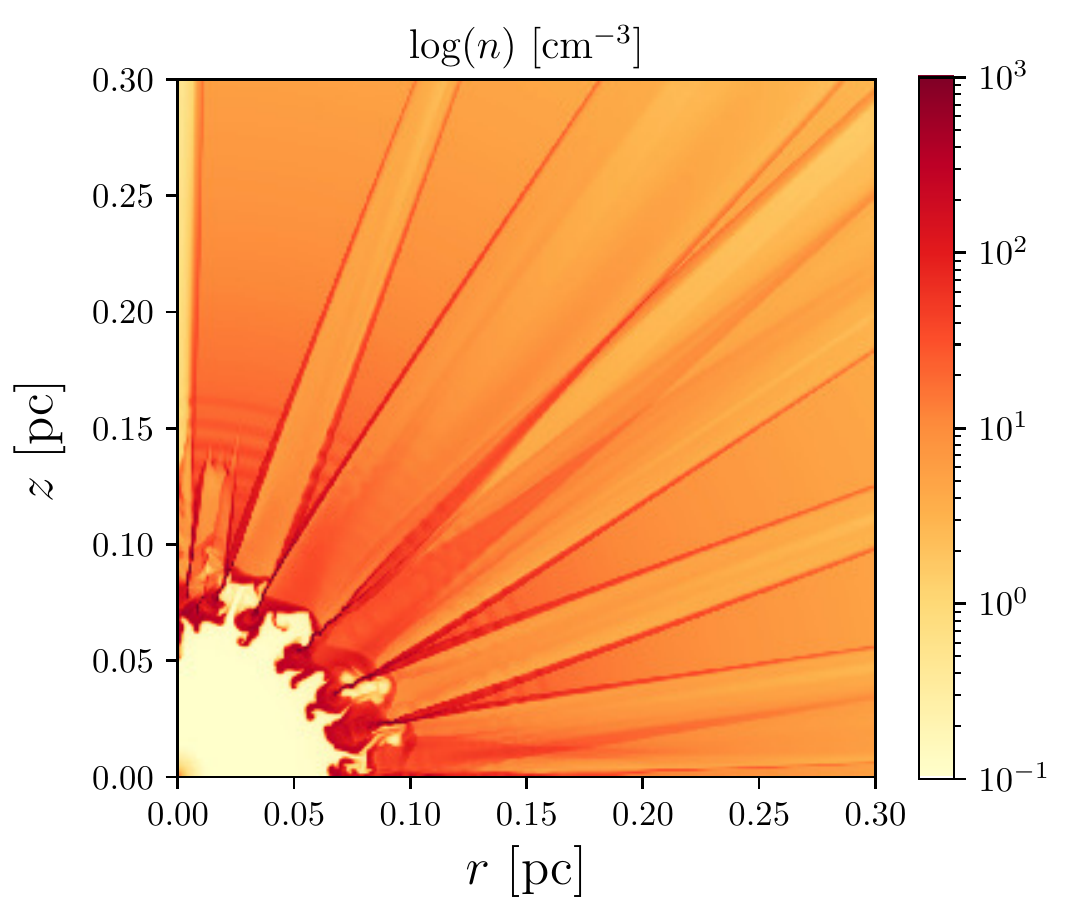}\\
  \vspace*{-0.2cm}
  \includegraphics[width=0.33\linewidth]{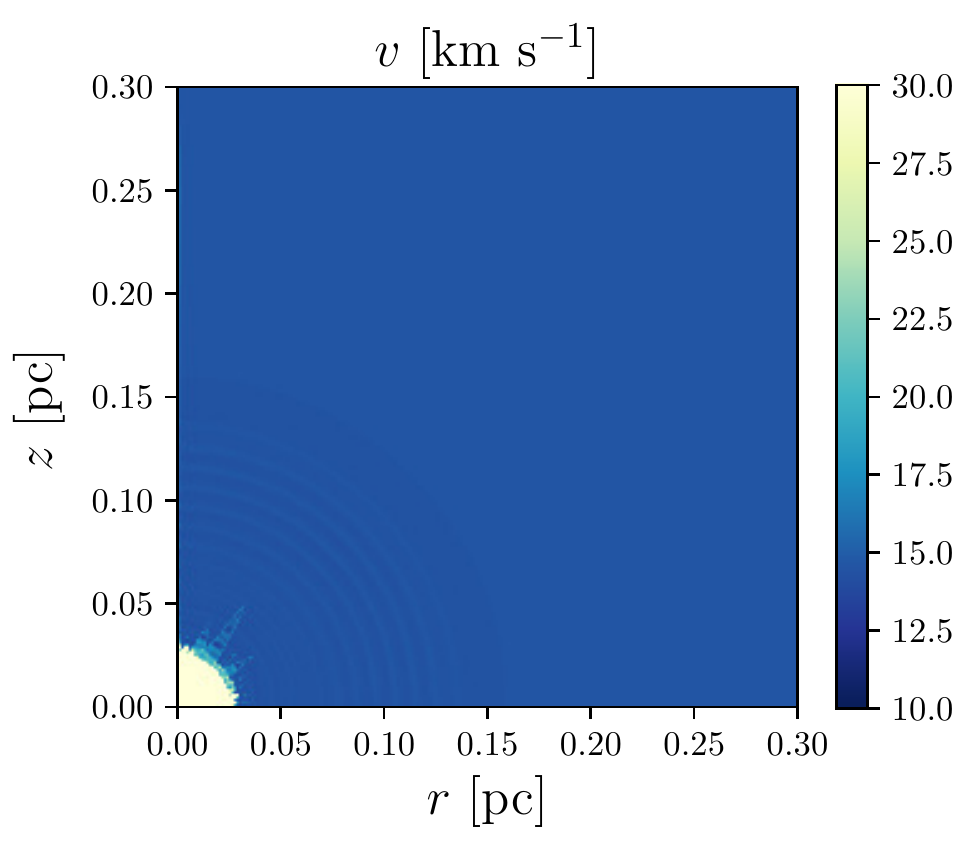}~
  \includegraphics[width=0.33\linewidth]{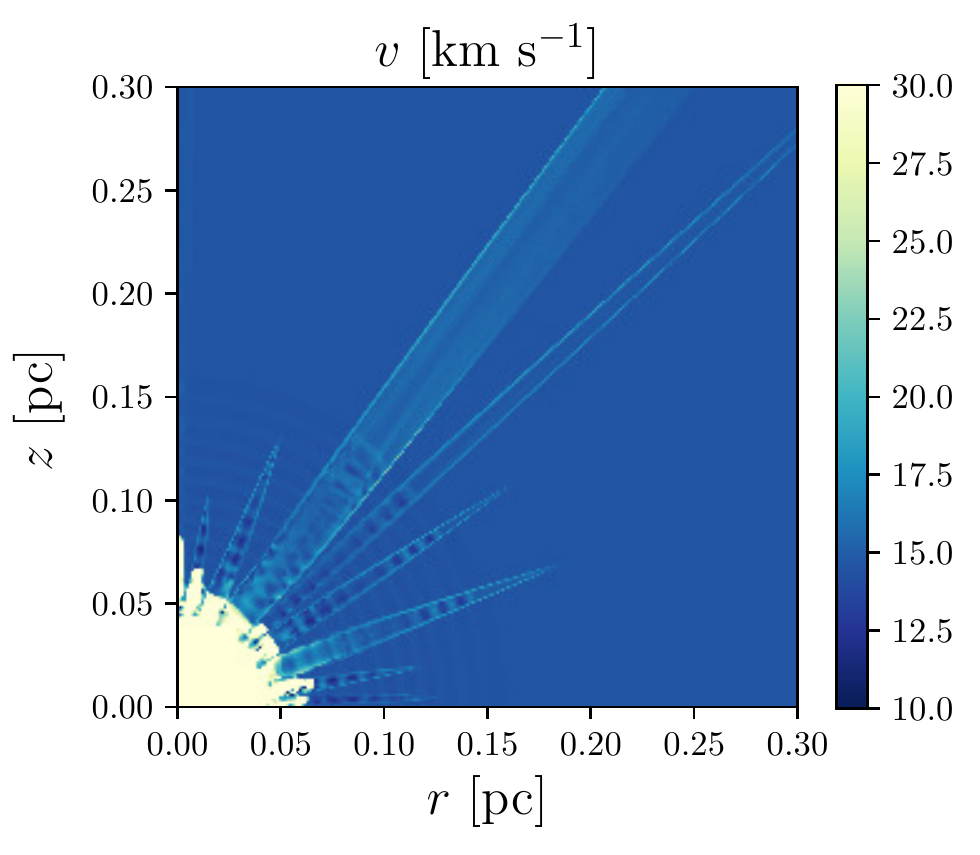}~
  \includegraphics[width=0.33\linewidth]{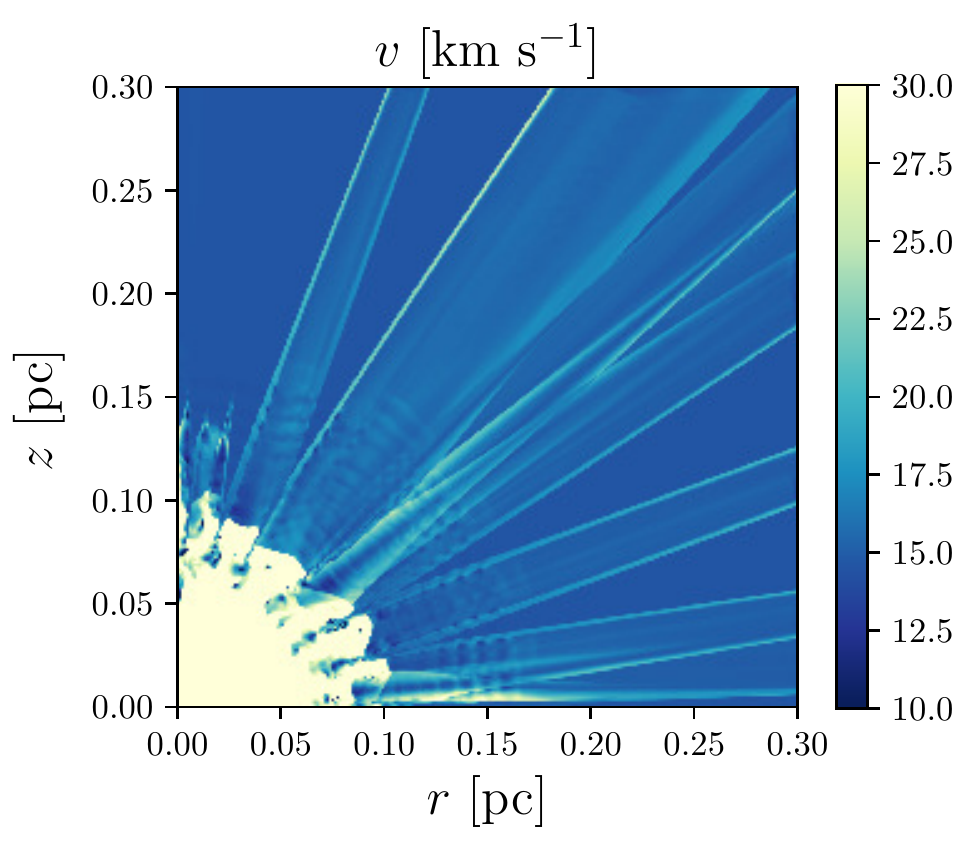}\\
  \vspace*{-0.2cm}
  \includegraphics[width=0.33\linewidth]{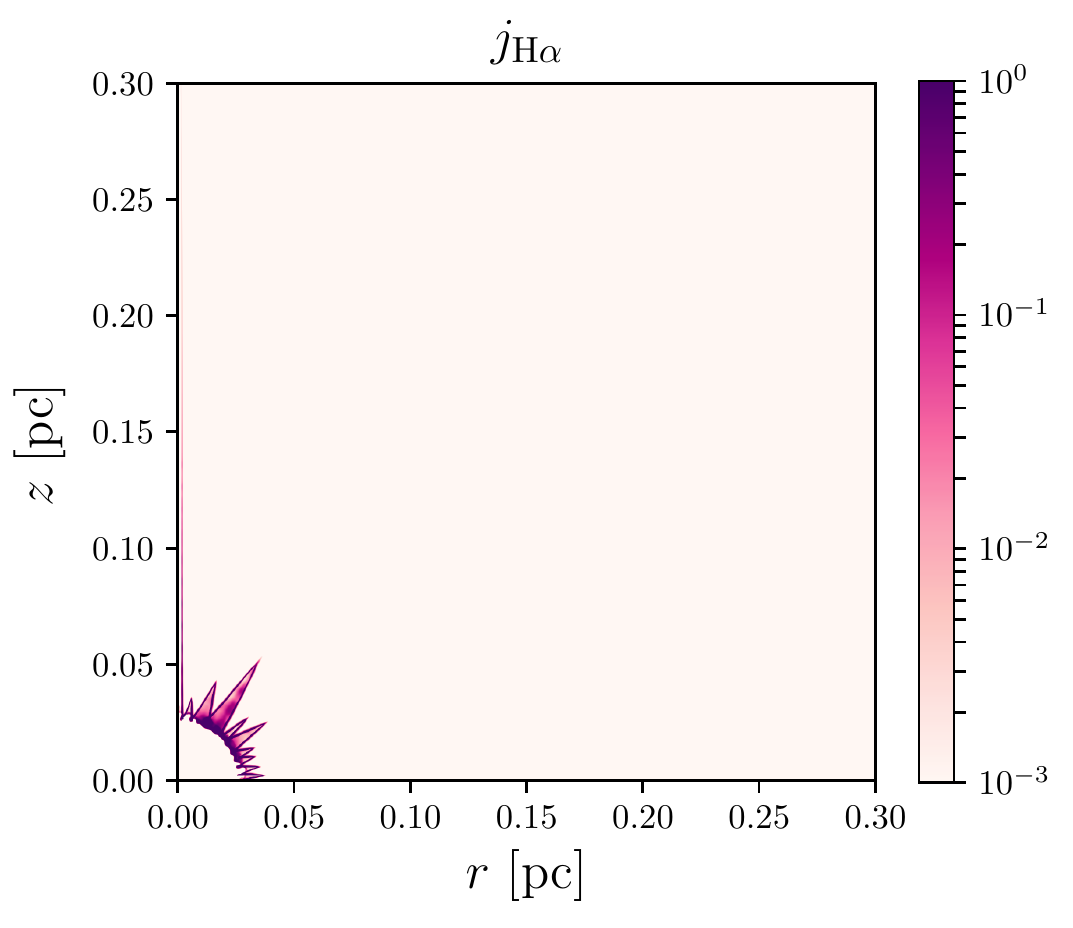}~
  \includegraphics[width=0.33\linewidth]{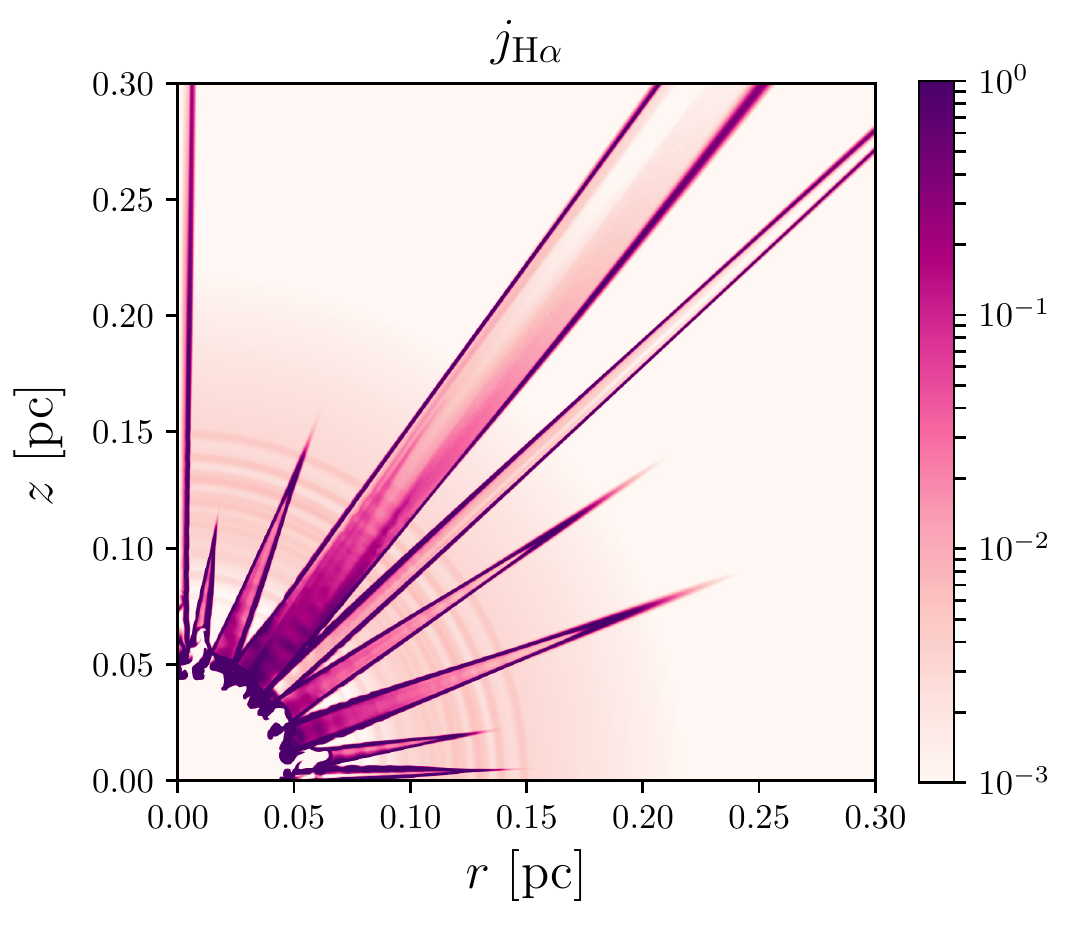}~
  \includegraphics[width=0.33\linewidth]{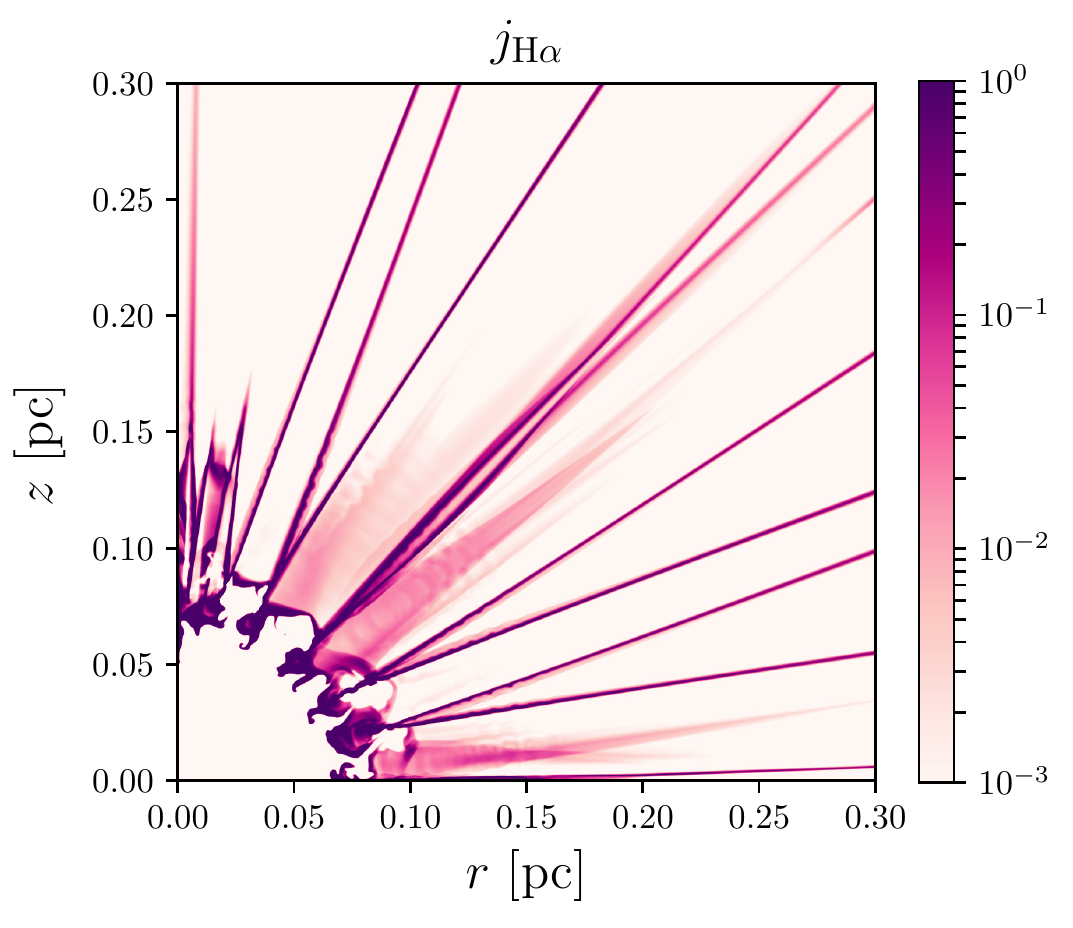}
  \caption{
Evolution with time of the total number density $n$ (top panels),
gas velocity $v$ (middle panels), and normalised H$\alpha$ volume
emissivity (bottom panels). 
Left, center, and middle columns correspond to 
2000, 4000, and 6000~yr of post-AGB evolution.
}
\label{fig:sim_2D}
\end{center}
\end{figure*}

In the simulations, the fast wind from the post-AGB phase expands
into the slow previously ejected AGB material, leading to the
creation of a two-shock pattern: the outer shocks sweeps and
compresses the AGB material, while the inner shocks thermalises
the fast wind creating a diffuse and hot inner bubble \citep{Dyson1997}.
Since the density is higher close to the star, the opacity is also
higher and the ionization front is initially trapped close to the
inner rim.  As the inner rim expands, the opacity drops, allowing the
photon flux to move outwards, but clumps formed at early times at the
wind-wind interaction layer due thin shell instabilities corrugate the
inner rim, producing variations in the opacity and shadowing
instabilities \citep[][]{Williams1999,Arthur2006}.

Figures~\ref{fig:sim_2D} shows three snapshots of the total number density
$n$, gas velocity $v$, and normalised H$\alpha$ volume emissivity
$j_{{\rm H}\alpha}$ derived from the 2D numerical simulations at times 2000,
4000, and 6000~yr to illustrate the effects that the variations in opacity 
produce in the early, mid-age and late evolution of external thin shells
of enhanced density.
We note that the formation of clumps in the wind-wind interaction at certain
angles is actually induced by numerical artifacts of the Cartesian grid used
in axisymmetric codes, as the spherically symmetric wind gas flows across
square cell boundaries.  
Whereas the spatial distribution of these clumps should be considered
non-relevant, their varying opacities and the effects of the shadowing
instabilities that they cause on the interaction of the ionizing flux
with the external rings are robust results of these simulations.
In early evolutionary phases ($t=2000$ yr), when the inner 
shell opacity is still high, the outer rings keep their initial 
structure.
Later on, when clumps form and the ionization is not isotropic any longer 
($=4000$ yr), streaks of ionized material and shadowing cones start
modifying the inner rim and the AGB material \citep[see][]{GS1999}, in
particular the rings.
Whilst the precise angular distribution of ionizing streaks and shadowing
cones is not indicative of their true spatial locations, their varying
intensities allow us to probe different degree of clumpiness in the
wind-wind interaction layer.
In late phases, when the inner rim becomes more unstable and
Rayleigh-Taylor instabilities start to dominate ($t=6000$ yr), 
resulting in dense, finger-like structures pointing inwards, 
initial rings are broken into independent arcs.
The dynamical effects induced by these processes also produce spatial
variations in the velocity field (middle panel of Figure~\ref{fig:sim_2D}), 
with independent arcs resulting from the break-up of the rings by the
ionization streaks expanding with different velocities.

The simulations presented above show that the effects of clumps formed 
at the wind-wind interaction zone trapping the ionizing flux during the
post-AGB phase can indeed distort coherent structures formed at the end
of the AGB, effectively splitting them apart into independent arcs rather
than completely erasing them as predicted in 1D simulations
\citep[e.g.,][]{Meijerink2003}.
Preliminary models for different initial masses of the progenitor
and properties of the mass-loss gasps during the AGB (mass loss
rate, duration, and period) are suggestive of a wealth of outcomes
and timescales. 
A comprehensive investigation of different initial
conditions will be subject of a subsequent study
(Toal\'a et al., in preparation, Paper~III).
The effects of anisotropic illumination and ionization patterns resulting
from shadowing instabilities are expected to dilute in 3D regular patterns,
but projection effects will be important.  
Furthermore, these disruptive processes will add to the noticeable spatial
and kinematic asymmetries of the 3D spiral structures proposed to arise in
the interaction of a binary system.
Modelling the evolution of 3D spiral patterns since their formation in the 
late AGB to the post-AGB phase, accounting for non-isotropic ionizing flux 
at that time, will be the topic of future investigations.

\section{Concluding Remarks}

We have investigated the detailed morphology and expansion velocity
of the spiral pattern around the AGB carbon-rich star AFGL\,3068
and the ring-like features around the PNe NGC\,6543, NGC\,7009 and
NGC\,7027 using high-quality multi-epoch \emph{HST} images to shed
light onto the possible evolutionary connection between these features.
AFGL\,3068, as some other AGB stars, shows a complete, regular
spiral structure around it, whereas the PNe here considered,
as all other proto-PNe and PNe, present incomplete ring-like
features and arcs.  
However, the spiral pattern around AFGL\,3068 and the arcs around these
three PNe have similar expansion velocities, linear sizes, and inter-lapse
times.
This agreement is supported by detailed studies of a number of AGB stars
and a statistical investigation of a large sample of proto-PNe and PNe
with similar features (Paper~I), suggesting that the ring-like features
in PNe may evolve from the regular patterns of AGB stars.
A statistical investigation of the kinematics of these
features is certainly needed to strenghten this idea.

Interestingly, the spiral pattern around AFGL\,3068 gets distorted
as material gets older and expands further out from the central star
as shown in recent hydrodynamical simulations \citep{Kim2019}. 
These effects are to be added to those resulting from the onset
of the post-AGB phase, including the expansion of the PN within
the AGB wind and the injection into it of an anisotropic ionizing
flux.  
We witness the interaction of the ring-like features around these
PNe with the nebular shells and collimated outflows, as they expand
and drive shocks that sweep up the material ejected during the AGB
phase.
We demonstrate using radiative-hydrodynamic simulations of PN
formation that instabilities in the wind-wind interaction region
produce anisotropies in the ionizing pattern of the material
ejected during the AGB phase.
This has important dynamical effects in the evolution of
regular patterns imprinted in the AGB material, which
get distorted and finally erased during the post-AGB
phase.

If this were the case for the ring-like features around NGC\,6543, NGC\,7009, 
and NGC\,7027, it would not be possible to unambiguosuly attribute them to 
previous spiral patterns.  
Interestingly, the inter-ring spacing at one side of each of these 
PNe is smaller and more coherent than at the other side, and the
rings are sharper.  
This asymmetry cannot be attributed to the compression of regular imprints in
the AGB wind as the PN moves through a dense interstellar medium, since these
three PNe are surrounded by much larger round haloes that do not reveal such
interactions \citep{MCW1989,MFG1998,NCM2003}.  
On the contrary, this asymmetry is naturally explained in the framework 
of a binary interaction producing spiral patterns with this imprint 
\citep{Kim2019}.  
On the assumption that the ring-like features around
PNe were once spiral patterns, then the time-lapse 
between these ring-like features is related to the 
orbital period of binary systems at their hearts.

\section*{Acknowledgements}

The authors acknowledge support from grants AYA
2014-57280-P and PGC2018-102184-B-I00, co-funded
with FEDER funds.  
JAT and MAG are funded by UNAM DGAPA PAPIIT project IA100318,
and LS by UNAM DGAPA PAPIIT project IN101819.
GRL acknowledges support from Fundaci\'on Marcos Moshinsky, CONACyT and
PRODEP (Mexico).
BB acknowledges support from program AR-14563 that was provided by NASA
through a grant from the Space Telescope Science Institute, which is
operated by the Association of Universities for Research in Astronomy,
Inc., under NASA contract NAS 5-26555.

This paper is based on observations made with the NASA/ESA
\emph{Hubble Space Telescope},
and obtained from the Hubble Legacy Archive, which is a collaboration between
the Space Telescope Science Institute (STScI/NASA), the Space Telescope
European Coordinating Facility (ST-ECF/ESA) and the Canadian Astronomy Data
Centre (CADC/NRC/CSA).
The large scale image of NGC\,7009 is based on observations collected at
the European Southern Observatory by the MUSE science verification team.  
This work has made extensive use of the NASA's Astrophysics Data System.


\end{document}